\documentclass[%
 reprint,
superscriptaddress,
eprint,
 amsmath,amssymb,
 aps,
pre,
]{revtex4-1}
\usepackage{graphicx}
\usepackage{dcolumn}
\usepackage[left=3cm,top=2.5cm,right=3cm,bottom=2.5cm]{geometry}
\usepackage{array}
\usepackage{color}
\usepackage{soul}
\usepackage{braket}
\usepackage{verbatim}
\usepackage{mathtools}
\usepackage{mathrsfs}
\usepackage{amsfonts}
\usepackage{physics}
\usepackage[english]{babel}
\usepackage{textcomp}

\usepackage{hyperref}
\hypersetup{pdfkeywords={Path integral, Effective conductivity}}
\usepackage{natbib}

\newcommand*\pd[1]{\mathop{\mathcal{D} #1}}

\makeatletter
\newcommand*{\Kbar}{}%
\DeclareRobustCommand*{\Kbar}{%
  \mathpalette\@Kbar{}%
}
\newcommand*{\@Kbar}[2]{%
  \sbox0{$#1\mathrm{K}\m@th$}%
  \sbox2{$#1K\m@th$}%
  \rlap{%
    \hbox to\wd2{%
      \hfill
      $\overline{%
        \vrule width -1pt height\ht0 %
        \kern\wd0 %
      }$%
    }%
  }%
  \copy2 %
}
\makeatother

\begin{document}


\title{Path Integral Renormalization of Flow through Random Porous Media}


\author{Umut C. \"Ozer}
\affiliation{The Blackett Laboratory, Imperial College London, London SW7 2AZ, United Kingdom}

\author{Peter R. King}
\affiliation{Department of Earth Science and Engineering, Imperial College London, London SW7 2BP, United Kingdom}

\author{Dimitri D. Vvedensky}
\affiliation{The Blackett Laboratory, Imperial College London, London SW7 2AZ, United Kingdom}

\begin{abstract}
 The path integral for Darcy's law with a stochastic conductivity, which characterizes flow through random porous media, is used as a basis for Wilson renormalization-group (RG) calculations in momentum space. A coarse graining procedure is implemented by integrating over infinitesimal shells of large momenta corresponding to the elimination of the small scale modes of the theory. The resulting one-loop $\beta$-functions are solved exactly to obtain an effective conductivity in a coarse grained theory over successively larger length scales. We first carry out a calculation with uncorrelated Gaussian conductivity fluctuations to illustrate the RG procedure before considering the effect of a finite correlation length of conductivity fluctuations. We conclude by discussing applications and extensions of our calculations, including comparisons with the numerical evaluation of path integrals, non-Gaussian fluctuations, and multiphase flow, for which the path integral formulation should prove particularly useful.
\end{abstract}

\maketitle

\section{Introduction}
\label{sec1}

Flow through porous media plays a pivotal role in many branches of science and engineering, including geologic CO$_2$ sequestration \cite{juanes12}, enhanced oil recovery \cite{orr84}, water infiltration into soil \cite{juanes08}, filtration \cite{zamani09}, as well as casting and solidification \cite{stefanescu02}.  Examples involving porous biological tissue \cite{khaled03} include the flow of oxygen through lungs \cite{miguel12}, the transport of cerebrospinal fluid through the brain \cite{penn07} and hemodynamics through blood vessels \cite{zunino10}.  Such widespread applications have stimulated the development of a broad range of theoretical and computational methods for describing flow through porous media, including the Navier--Stokes equation \cite{stanley99}, lattice Boltzmann methods \cite{chen98}, cellular automata \cite{rothman88}, percolation theory \cite{hunt14, king18}, and field theory \cite{king87}.

These methods vary substantially in their spatial resolution and the length scale over which they are valid, which is symptomatic of the inherent multiscale nature of flow through porous media \cite{sahimi93}. At the pore scale, fluid motion through the porous medium is obtained either directly from the Navier--Stokes equation, or indirectly from lattice Boltzmann simulations or cellular automata.  Although the microstructure of pores in natural media, such as rocks, can be measured with X-ray tomography \cite{blunt17}, pore-scale simulations are impractical for large regions because of their prohibitive computational requirements.  An alternative representation of the physics that enables large regions to be studied is based on coarse-graining the medium and flow over volumes large on the pore scale, but small on the scale of the system.  The porous medium can then be characterized by an effective correlated stochastic conductivity, whereupon flow can be calculated over macroscopic regions.

Determining the effects of pore structure on macroscopic flow remains a challenging theoretical problem.  In this paper, we restrict ourselves to a continuum description based on Darcy's law with a stochastic conductivity whose solution is represented as a path integral.  Applying the renormalization group (RG) to integrate infinitesimal shells in momentum space reduces the usual recursion relations to a set of coupled differential RG (DRG) equations for the trajectories in parameter space that provide an effective Darcy equation over successively larger length scales.  This enables the effect of conductivity correlations and the renormalization of conductivity distributions to be investigated.  The combination of the path integral solution to Darcy's law and the RG provides for a systematic methodology for a coarse-grained description of Darcy's equation with a random permeability that complements and, in some cases, extends previous RG calculations in real \cite{king89,morris90,king93,king02,neuweiler03,pancaldi07,pancaldi08,pancaldi09} and momentum \cite{teodorovich97,hristopulos99,sposito01,attinger03,eberhard04,hanasoge17} space.

The path integral for Darcy's law can be evaluated  numerically \cite{westbroek19a,westbroek19b} to obtain the pressure statistics of the flow for stochastic conductivities with various correlation lengths and boundary conditions.  These results provide benchmarks for the results reported here, but the analytic character of our RG analysis offers several advantages over the numerical approach: (i) different spatial dimensions are handled by the same basic procedure, whereas numerical evaluation necessitates different methodologies, (ii) changes to the initial distribution and correlation length of the conductivity along the RG trajectory determine the appropriate choices of these quantities at various length scales, and (iii) the extension to more complex settings, e.g., multiphase flow, can, with suitable modifications,  be carried out within the same framework.

Our paper is organized as follows: The path integral formulation for solutions to Darcy's law with a random permeability is derived in Sec.~\ref{sec2}. The RG transformations are then developed for the basic case of uncorrelated Gaussian permeability fluctuations in Sec.~\ref{sec3}, which concludes by considering integrals over infinitesimal momentum shells to derive the DRG equations. These are solved exactly in Sec.~\ref{sec:rg_trajectories_and_fixed_points} after the parameter trajectories and asymptotic behavior toward fixed points are determined. The generalization of our results to finite correlation lengths is described in Sec.~\ref{sec5}.  We summarize our results and discuss various applications and extensions in Sec.~\ref{sec6}. Detailed calculations and a Feynman diagrammatic analysis may be found in the appendices.

\section{Path Integral for Darcy's Law}
\label{sec2}

For the slow flow of incompressible viscous fluid, the average flow rate and the average pressure gradient are related by a source-free stochastic differential equation known as Darcy's law,
\begin{equation}
\label{eq:darcy's_law}
\div(K(\mathbf x) \grad \phi(\mathbf x)) = 0\, ,
\end{equation}
where $K(\mathbf x)$ is a random conductivity taken from a probability distribution function (PDF) $P[K]$.  In the case of flow through porous media, the hydraulic conductivity $K(\mathbf x) = k\rho g /\mu$ depends on the specific permeability $k$ of the medium, the viscosity $\mu$, and density $\rho$ of the fluid. The field $\phi(\mathbf x) = p/\rho + gz$ (energy per unit mass) is sometimes referred to as the piezometric head. Proposed as an empirical relation in the mid-1850s \cite{darcy56}, Darcy's law can be derived by homogenization of Stokes flow \cite{whitaker86}, and for flow through a dilute random array of fixed obstacles \cite{rubinstein87}.

Broadening the range of applicability of \eqref{eq:darcy's_law}, any solutions of Darcy's law are valid for systems governed by Ohm's, Fourier's, or Fick's laws involving a random, spatially varying electromagnetic permittivity $\varepsilon(\mathbf x)$, thermal conductivity $\kappa (\mathbf x)$, or diffusion constant $D(\mathbf x)$, respectively, in place of the hydraulic conductivity $K(\mathbf x)$.

\subsection{Path integral formulation}
\label{sec2.1}

Following the method of De Dominicis and Peliti \cite{dedominicis78}, we define a partition functional as a path integral over all realizations of $\phi$ and $K$ consistent with Darcy's law:
\begin{equation}
  Z_K = \int \pd{\phi} \delta[\div(K\grad\phi)]J\, .
\end{equation}
The functional generalization of the $\delta$-function enforces Darcy's law \eqref{eq:darcy's_law}. The Jacobian $J$ associated with the $\delta$-function is a constant because Darcy's law is linear.  This constant will cancel when we calculate averages and so will be neglected in what follows.

The $\delta$-functional can be written as a path integral through a functional Fourier transform:
\begin{align}
&\delta[\div(K\grad\phi)]\nonumber\\
&\propto \int \pd{\hat{\phi}} \exp[-\int \dd[d]{x} \hat{\phi} \div(K\grad\phi)]\, ,
\end{align}
where the response field $\hat{\phi}(\mathbf x)$ is purely imaginary. The action $S[\phi, \hat{\phi}]$ is found by bringing the partition functional into the form
\begin{equation}
  Z = \int \pd{\phi} \pd{\hat{\phi}} e^{-S_K[\phi, \hat{\phi}]}\, ,
  \label{eq2.4}
\end{equation}
where 
\begin{align}
S_K[\phi, \hat{\phi}] &= \int d^dx\, \hat{\phi} \div(K\grad\phi)\nonumber \\
   &= - \int d^dx\, K \grad\hat{\phi}\cdot\grad\phi\, .
   \label{eq2.5}
\end{align}
The action is obtained by integrating (\ref{eq2.4}) over all realizations of the conductivity,
\begin{equation} \label{eq:integral_over_S_K}
  e^{-S[\phi, \hat{\phi}]} = \int \pd{K} P[K] e^{-S_K[\phi, \hat{\phi}]}\, .
\end{equation}

Expectation values of products of the fields $\phi$ and $\hat{\phi}$ are computed by introducing the auxiliary currents $J$ and $\hat J$ into the partition functional,
\begin{equation}
  Z[J, \hat J] = \int \pd{\phi} \pd{\hat{\phi}} e^{-S + \int d^dx\,(\hat J \phi + J \hat{\phi})}\, .
  \label{eq2.7}
\end{equation}
In particular, the connected correlation functions are calculated as
\begin{align}
&\langle \phi_1 \cdots \phi_n \hat{\phi}_1 \cdots\hat{\phi}_m \rangle_c\nonumber\\
  &= \frac{\int \pd{\phi} \pd{\hat{\phi}} \phi_1 \cdots \phi_n \hat{\phi}_1 \cdots \hat{\phi}_m e^{-S}}{\int \pd{\phi} \pd{\hat{\phi}} e^{-S}}\nonumber\\
 &= \left. \frac{\delta^{n + m} \ln Z}{\delta \hat J_1 \cdots \delta \hat J_n \delta J_1 \cdots \delta J_m} \right\rvert_{J = \hat J = 0}.
 \label{eq2.8}
\end{align}
Any constant prefactors appearing in $Z$ will therefore cancel and are neglected in this paper.

\subsection{Bare Propagator}
\label{sec2.2}

The bare propagator is determined from the action by using a constant conductivity equal to the mean of the stochastic conductivity:
\begin{equation}
  \label{eq:Kbar}
  \Kbar = \frac{\int \dd[d]{x} K(\vb{x})}{\int \dd[d]{x}}\, .
\end{equation}
The corresponding action $S_0$ is obtained by replacing $K({\bf x})$ by $\Kbar$ in (\ref{eq2.5}), followed by a Fourier transform:
\begin{equation}
 S_0[\phi, \hat{\phi}]  = \int \frac{d^dk}{(2\pi)^d} k^2 \Kbar \hat{\phi}_{-\vb k} \phi_{\vb k}\, ,
 \label{eq9}
\end{equation}
in which $\phi_{\vb k}$ and $\hat{\phi}_{\vb k}$ are the Fourier modes of $\phi_{\vb{x}}$ and $\hat{\phi}_{\vb{x}}$, respectively, with wave vector $\vb k$, defined as:
\begin{align}
\phi_{\vb k} = \int d^dx\, e^{-i\vb k \vdot \vb x} \phi_{\vb x}\, ,\\
\hat{\phi}_{\vb k} = \int d^dx\, e^{-i\vb k \vdot \vb x} \hat{\phi}_{\vb x}\, .
\end{align}
The partition function that generates the propagator is then obtained as in (\ref{eq2.7}) by introducing auxiliary currents $J$ and $\hat J$ and carrying out the resulting Gaussian functional integration over $\phi$ and $\hat\phi$:
\begin{widetext}
  \begin{align}
    Z[J, \hat J] &= \int \pd{\phi}\pd{\hat{\phi}} \exp[\int \frac{d^dk}{(2\pi)^d}\left( k^2 \Kbar \hat{\phi}_{-\vb k} \phi_{\vb k} + \hat J_{-\vb k}\phi_{\vb k} + \hat{\phi}_{-\vb k}J_{\vb k} \right)]\nonumber\\
    &\propto \exp[\int \frac{d^d{k}}{(2\pi)^d}\frac{d^d{k'}}{(2\pi)^d} \hat J_{-\vb k} D(\vb k - \vb k') J_{\vb k'}]\, ,
  \end{align}
\end{widetext}
where the Feynman propagator $D(\vb{k} - \vb{k}')$ satisfies
\begin{equation}
  k^2 \Kbar D(\vb k - \vb k') = (2\pi)^d \delta^d(\vb k - \vb k')\, .
\end{equation}
After integrating over ${\bf k}'$, we arrive at the generating functional
\begin{equation}
  Z[J, \hat J] \propto \exp(\int \frac{\dd[d]{k}}{(2\pi)^d} G_0(k) \hat J_{-\vb k} J_{\vb k}),
\end{equation}
where the bare propagator is
\begin{equation}
  G_0(k) = \frac{1}{k^2 \Kbar}\, .
\end{equation}
Finally, the two-point response function is obtained by taking the functional derivatives
\begin{equation}
  \begin{split}
    \langle \phi_{\vb k} \hat{\phi}_{\vb k'} \rangle_{0} &= \left.\frac{\delta^2 \ln Z}{\delta \hat J_{\vb k} \delta J_{\vb k'}} \right\rvert_{J = \hat J = 0} \\
    \noalign{\vskip3pt}
    &= (2\pi)^d \delta^d(\vb k' + \vb k) G_0(k)\, .
    \label{eq2.15}
  \end{split}
\end{equation}
The notation $\langle \cdots \rangle_0$ signifies the path integral over the Gaussian ensemble $e^{-S_0}$ for constant conductivity. The bare propagator is essentially all we need to use the path integral formalism because Wick's theorem tells us that any other correlation function can be decomposed into products of two-point correlators. Furthermore, any non-Gaussian terms $S_I = S - S_0$ in the action can be dealt with perturbatively by computing the various correlation functions arising from the expansion of the exponential $e^{-S_I}$. This will be explained in Sec.~\ref{sec3}.

\subsection{Gaussian Conductivity Fluctuations}%
\label{sec2.3}

One of the simplest distributions for the conductivity is a Gaussian. To analyze this case, the conductivity can be written in terms of fluctuations $\widetilde{K}$ about the mean conductivity $\Kbar$ defined in \eqref{eq:Kbar}:
\begin{equation}
  K(\vb x) = \Kbar + \widetilde{K}(\vb x) \, .
\label{eq2.16}
\end{equation}
This splitting allows us to separate the action into the free part $S_0$ in \eqref{eq9} and an interaction term that encodes the conductivity fluctuations:
\begin{align}
    S_K = &-\int d^d{x} \Kbar \grad\hat{\phi}_{\vb x} \vdot \grad\phi_{\vb x}\nonumber\\
    &-\int d^d{x} \widetilde K_{\vb x} \grad\hat{\phi}_{\vb x} \vdot \grad\phi_{\vb x}\, .
\end{align}
We consider the case where the fluctuations $\widetilde{K}$ are taken from a Gaussian PDF of the form
\begin{equation}
  P[\widetilde K] \propto \exp(-\frac{1}{2} \int d^d{x} d^d{y} \widetilde K_{\mathbf x} C^{-1}_{\vb{x}, \vb{y}} \widetilde K_{\vb{y}})\, ,
\end{equation}
where $C^{-1}_{\vb{x}, \vb{y}} = C^{-1}(\vb{x}, \vb{y})$ satisfies
\begin{equation}
  \int \dd[d]{y} C^{-1}(\vb{x}, \vb{y}) C(\vb{y}, \vb{z}) = \delta^{d}(\mathbf x - \mathbf z)\, ,
\end{equation}
and the two-point correlation function is
\begin{equation}
  C(\vb{x}, \vb{y}) = \langle \widetilde K(\vb x) \widetilde K(\vb y) \rangle_K \, ,
\end{equation}
in which the angular brackets $\langle \cdots \rangle_K$ denote the path integral weighted by $P[K]$.

By using \eqref{eq:integral_over_S_K} and \eqref{eq2.16}, the action $S$ is  found by  integrating over $\pd{K} P[K] = \pd{\widetilde K} P[\widetilde K]$:
\begin{widetext}
  \begin{align}
    \label{eq:action_as_fn_of_C}
    e^{-S} &=\int \pd{K} P[K] \exp[\int d^d{x} K_{\vb x} (\nabla \hat{\phi} \cdot \nabla \phi)_{\vb x}]\nonumber\\
    &= e^{-S_0} \exp[-\frac{1}{2} \int d^d{x}\, d^d{y} (\nabla \hat{\phi} \cdot \nabla \phi)_{\vb x} C_{\vb x, \vb y} (\nabla \hat{\phi} \cdot \nabla \phi)_{\vb y}]\, .
  \end{align}
\end{widetext}
The simplest choice is the delta correlation,
\begin{equation}
  \label{eq:delta_correlation}
  C(\vb x, \vb y) = \sigma^2 (2\pi \Lambda^{-2})^{d/2} \delta^d(\vb x - \vb y),
\end{equation}
where $\sigma^2$ is the variance and the quantity $\Lambda^{-1}$ is the smallest length scale over which Darcy's law remains valid. 
Being the only length scale in the theory, the factor $\Lambda^{-d}$ provides dimensional consistency. The normalisation $(2\pi)^{{d}/{2}}$ is chosen in anticipation of the result \eqref{eq:vanishing_cor_len} of Sec.~\ref{sec5}, which shows that (\ref{eq:delta_correlation}) arises naturally in the limit of vanishing correlation length. Finally, we note that the inverse length $\Lambda$ is the highest wavenumber in the system and can therefore be interpreted as the ultraviolet (UV) cutoff of our theory. 
With (\ref{eq:delta_correlation}), the action is
\begin{equation} \label{eq:action_gaussian_delta}
  S[\phi, \hat{\phi}] = \int d^d{x} [-\Kbar \grad\hat{\phi}\vdot\grad\phi + g(\grad\hat{\phi}\vdot\grad\phi)^2],
\end{equation}
where the coupling constant $g$, defined as
\begin{equation}
  g = \frac{1}{2} \sigma^2 (2\pi \Lambda^{-2})^{d/2}\, ,
  \label{eq24}
\end{equation}
measures the variance $\sigma^2$ of fluctuations in the $d$-dimensional system of length scale $\Lambda^{-1}$.

\section{Coarse Graining}
\label{sec3}

Starting from an action that describes a system up to a UV cutoff given by the wavenumber $\Lambda$, we implement the Wilson RG transformation \cite{wilson74,goldenfeld92} in three steps:~(i) the Fourier modes of the fields are separated into high and low wavenumber components, (ii) the high wavenumber components, which represent the small scale fluctuations, are integrated out. The terms in the action beyond quadratic order in the fields, which encode information about the conductivity fluctuations, are expanded in a perturbation series, and (iii) the effective action obtained by integrating over the small-scale fluctuations is rescaled to restore the UV cutoff $\Lambda$. This enables a comparison of the transformed effective action with the initial action, resulting in RG recursion relations, which determine effective parameters for the coarse grained system. 

If the fluctuations are integrated over infinitesimal shells of high wavenumbers, the recursion relations reduce to DRG equations \cite{chang92}. In that case, the wealth of techniques for solving coupled autonomous nonlinear differential equations can applied to solving these equations.  We will formulate the DRG equations to one-loop order.  The steps of the RG are carried out in the following sections.

\subsection{Separation of Fourier Modes}%
\label{sub:separation_of_fourier_modes}


The first step of the RG transformations is to partition the fields into high and low wavenumber components,
\begin{equation}
  \phi_{\vb k} = \phi^{+}_{\vb k} + \phi^-_{\vb k}, 
  \qquad
  \hat{\phi}_{\vb k} = \hat{\phi}^{+}_{\vb k} + \hat{\phi}^-_{\vb k},
  \label{eq3.24}
\end{equation}
where the split is carried out at some fraction $\Lambda/\zeta$ of the UV cutoff. The low wavenumber, long wavelength fluctuations are given by
\begin{equation} \label{eq:low_modes}
  \phi^-_{\vb k} =
  \begin{cases}
    \phi_{\vb k}, & \abs{\vb k} < \Lambda / \zeta \\
    0, & \abs{\vb k} > \Lambda / \zeta
  \end{cases},
\end{equation}
while the high wavenumber fluctuations which we want to integrate are given by
\begin{equation} \label{eq:high_modes}
  \phi^+_{\vb k} =
  \begin{cases}
    0, & \abs{\vb k} < \Lambda / \zeta \\
    \phi_{\vb k}, & \Lambda > \abs{\vb k} > \Lambda / \zeta,
  \end{cases}
\end{equation}
and similarly for the response field $\hat{\phi}^+_{\vb k}$ and $\hat{\phi}^-_{\vb k}$.

The action (\ref{eq:action_gaussian_delta}) obtained for Gaussian conductivity fluctuations separates into the sum of two terms:~$S_0$, a term which is quadratic in the fields and contains only the mean conductivity $\Kbar$, and a nonlinear `interaction' term $S_I$ with quartic products of the fields obtained by including conductivity functions $\widetilde{K}$ about the mean.  The Fourier representation of $S_0$ is given in \eqref{eq9}, whereas $S_I$ can be written
\begin{align}
  \label{eq:S_I_fourier_zero_corlen}
  &S_I[\phi, \hat{\phi}] = \int \dd[d]{x} g(\grad\hat{\phi}\vdot\grad\phi)^2\nonumber\\
  \begin{split}
    &= \int \left[ \prod_{i=1}^4 \frac{\dd[d]{k_{i}}}{(2\pi)^d} \right] 
    \biggl\{g (\vb k_1 \vdot \vb k_2) (\vb k_3 \vdot \vb k_4) \\
    &\qquad \times \hat{\phi}_{\vb k_1} \phi_{\vb k_2} \hat{\phi}_{\vb k_3} \phi_{\vb k_4}
    (2\pi)^d \delta^d \left( \sum_{i=1}^4 \vb k_1 \right)\biggr\} ,
  \end{split}
\end{align}
The definitions \eqref{eq:low_modes} and \eqref{eq:high_modes} ensure that $S_0$ does not mix high and low wavenumber modes; any cross-terms $S_0[\phi^\pm, \hat{\phi}^\mp]$ vanish.  However, the nonlinear term $S_I$ does mix high and low wavenumber fields. Therefore, under a separation of Fourier modes, the action $S = S_0 + S_I$ decomposes as
\begin{align}
    S[\phi, \hat{\phi}] &= S_0[\phi^-, \hat{\phi}^-] + S_0[\phi^+, \hat{\phi}^+] \nonumber\\
    &\quad+ S_I[\phi^-, \hat{\phi}^+; \phi^+, \hat{\phi}^-].
\end{align}

\subsection{Integrating Over High wavenumbers}%
\label{sub:integrating_out_high_wave_numbers}

After separating the high and low wavenumber fields, the path integral measure splits as
\begin{equation}
  \mathcal{D}\phi = \mathcal{D}\phi^- \mathcal{D}\phi^+, \qquad
  \mathcal{D}\hat{\phi} = \mathcal{D}\hat{\phi}^- \mathcal{D}\hat{\phi}^+,
\end{equation}
which enables the integration over the large wavenumber modes in the partition function:
\begin{widetext}
\begin{equation} 
\label{eq31}
  Z = \int \pd{\phi^-}\pd{\hat{\phi}^-} \biggl\{ e^{-S_0[\phi^-, \hat{\phi}^-]} 
  \int \pd{\phi^+}\pd{\hat{\phi}^+} e^{-S_0[\phi^+, \hat{\phi}^+]} e^{-S_I[\phi^-, \hat{\phi}^+; \phi^+, \hat{\phi}^-]} \biggr\}.
\end{equation}
\end{widetext}
Writing the partition function as an integral over only the low wavenumber fields,
\begin{equation} \label{eq32}
  Z = \int \pd{\phi^-}\pd {\hat{\phi}^-} e^{-S'[\phi^-, \hat{\phi}^-]},
\end{equation}
defines the Wilsonian effective action $S'[\phi, \hat{\phi}]$, which describes the low wavenumber, large scale theory.
Comparison of \eqref{eq31} and \eqref{eq32} yields
\begin{equation}
  e^{-S'[\phi^-, \hat \phi^{-}]} = e^{-S_0[\phi^-, \hat{\phi}^-]} \langle e^{-S_I[\phi^{-}, \hat \phi^{+}; \phi^{+}, \hat \phi^{-}]} \rangle_{+},
\end{equation}
where $\langle \cdots \rangle_{+}$ denotes the integral over the high wavenumber fields with the Gaussian weight $e^{-S_0[\phi^+, \hat{\phi}^+]}$.
Taking logarithms of both sides yields an equation for the effective action:
\begin{equation}
  \label{eq:effective-action}
  S'[\phi^-, \hat \phi^-] = S_0[\phi^-, \hat{\phi}^-] - \ln \langle e^{-S_I[\phi^{-}, \hat \phi^{+}; \phi^{+}, \hat \phi^{-}]}\rangle_{+}.
\end{equation}
The integral over the fluctuation term $S_I$ can be treated perturbatively by assuming the variance of fluctuations $S_I \sim g \sim \sigma^2$ to be small. In this case, we can expand $e^{x}$ and then $\ln( 1+x )$ to obtain
\begin{align}\label{eq2.35}
  \ln \langle e^{-S_I} \rangle_{+} &= \ln \langle 1 - S_I + \frac{1}{2} S_I^2 - \dots \rangle_+ \nonumber \\
  &= -\langle S_I \rangle_{+} + \frac{1}{2} [\langle S_I^2 \rangle_{+} - \langle S_I \rangle^2_+] + \dots \nonumber \\
  &= \sum_{n=1}^{\infty} (-1)^n \langle S_I^n \rangle_{c, +}.
\end{align}
The final line indicates that the $n$\textsuperscript{th} term in the perturbation series is given by the $n$\textsuperscript{th} cumulant, or connected correlation function, of $-S_I$.

\subsubsection{Order g}%
\label{sub:order_g}

We will calculate the first two contributing terms using Wick's theorem. Appendix \ref{apx:diagrams} develops this calculation by using Feynman diagrams.

To order $O(g)$ the effective action is
\begin{align}
  &S'[\phi^-, \hat{\phi}^-] = S_0[\phi^-, \hat{\phi}^-] \nonumber\\ 
  &\quad+ \langle S_I[\phi^-, \hat{\phi}^+; \phi^+, \hat{\phi}^-] \rangle_{+} + O(g^2).
\end{align}
Since the form of $S_0$ is known, only the path integral $\langle S_I \rangle_+$ needs to be calculated. The separation (\ref{eq3.24})-(\ref{eq:high_modes}) into high and low wavenumber fields splits the product $\hat{\phi}_{\vb k_1}\phi_{\vb k_2}\hat{\phi}_{\vb k_3}\phi_{\vb k_4}$ in the Fourier expansion of $S_I$ into $2^4 = 16$ terms, which accounts for all combinations of $+$ and $-$ fields. Of these, we need to consider only terms containing an even number of $+$ fields, as integrals involving an odd number of these fields will vanish under the Gaussian ensemble. The term $\hat{\phi}^-\phi^-\hat{\phi}^-\phi^-$, containing only low wavenumber fields, is unaffected by the integral and simply adds the term $S_I[\phi^-, \hat{\phi}^-]$ to the effective action. At the other extreme, integrating over the purely high wavenumber term $\hat{\phi}^+\phi^+\hat{\phi}^+\phi^+$ yields an additive constant term to the effective action. This constant affects only the normalization of the partition function, which, as indicated in (\ref{eq2.8}), cancels when calculating averages and is, therefore, unimportant for the large-scale physics of the system.

The foregoing discussion indicates that the only types of terms that need to be calculated involve combinations of any two high and any two low wavenumber fields. Since any term of the form $\langle \cdots \rangle_+$ is a path integral over the Gaussian ensemble $e^{-S_0[\phi^+, \hat{\phi}^+]}$, we can invoke our result for the bare propagator in Sec.~\ref{sec2.2}. Therefore, only the two-point response function $\langle \hat{\phi}^+ \phi^+ \rangle_{+}$ will be nonzero, and we obtain, from (\ref{eq2.8}),
\begin{equation}
  \langle \hat{\phi}^+_{\vb k'} \phi^+_{\vb k} \rangle_{+} = (2\pi)^d \delta^d (\vb k' + \vb k) G_0(k).
\end{equation}
Integrals over the other combinations of two $+$ fields, namely $\langle \phi^+ \phi^+ \rangle_+$ and $\langle \hat{\phi}^+ \hat{\phi}^+ \rangle_{+}$, will both vanish.
The first contribution to the effective action is thus
\begin{widetext}
\begin{equation}
  \langle S_I \rangle_+ = 4 g \int \left[\prod_{i=1}^4 \frac{\dd[d]{k_i}}{(2 \pi)^d}\right] 
  (\vb k_1 \cdot \vb k_2)(\vb k_3 \cdot \vb k_4) 
  (2\pi)^d \delta^d\left( \sum_{i=i}^{4} \vb k_i \right)
  \hat{\phi}^-_{\vb k_1} \phi^-_{\vb k_2} 
  \underbrace{(2\pi)^d \delta^d (\vb k_3 + \vb k_4) G_0(k_4)}_{\langle \hat{\phi}^+_{\vb k_3} \phi^+_{\vb k_4}\rangle_+},
  \label{eq3.40}
\end{equation}
\end{widetext}
where the symmetry factor in front of the integral accounts for $4$ identical contributions that differ only by a relabeling of the wave vectors $\vb k_1 \leftrightarrow \vb k_2$ or $\vb k_3 \leftrightarrow \vb k_4$.
Performing the integrals over $\dd[d]{k_3}$ and then $\dd[d]{k_1}$ uses the two $\delta$-functions to first replace $\vb k_3 = -\vb k_4$ and then $\vb k_1 = -\vb k_2$. Finally, by relabeling ${\vb k_2 \rightarrow \vb k}$ and $\vb k_4 \rightarrow \vb q$, we arrive at 
\begin{equation}
  \label{eq:order_g_wicks_theorem}
  \langle S_I \rangle_+ = 4 \frac{g }{\Kbar} \int_0^{\Lambda/\zeta} \frac{\dd[d]{k}}{(2\pi)^d} k^2 \hat{\phi}^-_{-\vb k} \phi^-_{\vb k} \int_{\Lambda/\zeta}^\Lambda \frac{\dd[d]{q}}{(2\pi)^d}.
\end{equation}
Including $S_I[\phi^-, \hat{\phi}^-]$, which is unaffected by coarse-graining, the effective action $S'$ is, to $O(g)$, given by
\begin{align}
  S'[\phi^-, \hat{\phi}^-] &\approx -\int_0^{\Lambda / \zeta} \frac{\dd[d]{k}}{(2\pi)^d} k^2 \Kbar^\prime\hat{\phi}^-_{-\vb k} \phi^-_{\vb k}\nonumber\\
  &\quad+ S_I[\phi^-, \hat{\phi}^-].
\end{align}
Comparison between this action and the original action (\ref{eq:action_gaussian_delta}) suggests a renormalized average conductivity
\begin{equation}
  \Kbar^\prime = \Kbar - 4 \frac{g }{\Kbar} \int_{\Lambda/\zeta}^\Lambda \frac{\dd[d]{q}}{(2\pi)^d}.
  \label{eq2.42}
\end{equation}
However, to properly compare these actions, we must rescale the wave vectors to restore the initial integration limits. This will be done in Sec.~\ref{sub:rescaling_and_rg_equations}.

\subsubsection{Order \texorpdfstring{$g^2$}{g squared}}%
\label{sub:order_g_2}

Referring to (\ref{eq2.35}), the contribution to the effective action to $O(g^2)$  is obtained by computing the path integrals within
\begin{equation}
  \frac{1}{2}[\langle S_I^2 \rangle_+ - \langle S_I \rangle^2_+].
  \label{eq44a}
\end{equation}
The term $\langle S_I \rangle_+$ was calculated in (\ref{eq3.40}).  The contribution of $\langle S_I \rangle^2_+$ to the effective action will be discussed below. In real-space, $S_I^2$ is obtained from (\ref{eq:S_I_fourier_zero_corlen}) as
\begin{equation}
  S_I^2 = g^2 \int \dd[d]{x} \dd[d]{y} (\grad\hat{\phi}\vdot\grad\phi)^2_{\vb x}(\grad\hat{\phi}\vdot\grad\phi)^2_{\vb y}.
\end{equation}
Splitting the fields into high and low wavenumber components gives $2^8=256$ terms. However, we will only need to look at a subset of these terms. 
In particular, we focus on the terms whose Fourier space representation is of the form:
\begin{widetext}
  \begin{align} \label{eq:form_of_S_I^2_terms}
    \frac{1}{2} \langle S_I^2 \rangle_{+} 
    &\sim \frac{1}{2} (2^2)^2 g^2 \int_0^{\Lambda/\zeta} \left[\prod_{i=1}^4\frac{\dd[d]{k_i}}{(2 \pi)^d}\right] (\vb k_1 \cdot \vb k_2) (\vb k_3 \cdot \vb k_4) \hat{\phi}^-_{\vb k_1} \phi^-_{\vb k_2} \hat{\phi}^-_{\vb k_3} \phi^-_{\vb k_4}\nonumber \\
    &\times \int_{\Lambda/\zeta}^\Lambda \left[\prod^4_{i=1}\frac{\dd[d]{q_i}}{(2 \pi)^d}\right] (\vb q_1 \cdot \vb q_2) (\vb q_3 \cdot \vb q_4) \langle \hat{\phi}^+_{\vb q_1} \phi^+_{\vb q_2} \hat{\phi}^+_{\vb q_3} \phi^+_{\vb q_4}\rangle_+ \nonumber\\
    \noalign{\vskip6pt}
    &\times (2 \pi)^d \delta^d(\vb k_1 + \vb k_2 + \vb q_1 + \vb q_2) (2 \pi)^d \delta^d(\vb k_3 + \vb k_4 + \vb q_3 + \vb q_4).
\end{align}
\end{widetext}
After separating all fields into high and low wavenumbers, we choose a pair of conjugate fields $\hat{\phi}^-\phi^-$ from $(\grad\hat{\phi}\vdot\grad\phi)^2_{\vb x}$ and another pair from $(\grad\hat{\phi}\vdot\grad\phi)^2_{\vb y}$. The combinatoric factor $(2^2)^2$ reflects this choice. 
The labels $\vb k_i$ indicate low wavenumbers, and $\vb q_i$ high wavenumbers, which are to be integrated out. 
Two $\delta$-functions arise when performing the integrals over $\dd[d]{x}$ and $\dd[d]{y}$. Their arguments are given by specific combinations of wave vectors, which indicate that we chose a conjugate pair from each integral.
Finally, the factors $\vb k_i \vdot \vb k_j$ and $\vb q_i \vdot \vb q_j$ arise from the real space gradients. 
Wick's (or Isserlis') theorem \footnote{For Gaussian random variables, this result was established by Isserlis \cite{isserlis18}. In the context of quantum field theory, the analogous result is Wick's theorem \cite{wick50}.} stipulates that the term $\langle \hat{\phi}^+_{\vb q_1} \phi^+_{\vb q_2} \hat{\phi}^+_{\vb q_3} \phi^+_{\vb q_4} \rangle_+$ can be decomposed into a sum of all products of two-point correlators:
\begin{align} \label{eq44}
    &\langle \hat{\phi}^+_{\vb q_1} \phi^+_{\vb q_2} \hat{\phi}^+_{\vb q_3} \phi^+_{\vb q_4} \rangle_+
    = \langle\hat{\phi}^+_{\vb q_1} \phi^+_{\vb q_2}\rangle_+ \langle\hat{\phi}^+_{\vb q_3} \phi^+_{\vb q_4}\rangle_+ \nonumber\\
    &+ \langle\hat{\phi}^+_{\vb q_1} \hat{\phi}^+_{\vb q_3}\rangle_+ \langle\phi^+_{\vb q_2} \phi^+_{\vb q_4}\rangle_+ 
    + \langle\hat{\phi}^+_{\vb q_1} \phi^+_{\vb q_4}\rangle_+ \langle\hat{\phi}^+_{\vb q_2} \phi^+_{\vb q_3}\rangle_+.
\end{align}
Again, the only non-zero terms are those built from the propagator $\langle \hat{\phi}^+ \phi^+ \rangle_+$, while correlators of the same field type, such as $\langle \phi^+ \phi^+ \rangle_+$ or $\langle \hat{\phi}^+ \hat{\phi}^+ \rangle_+$, vanish. Hence, the second term on the right-hand side of \eqref{eq44} vanishes. 

On the other hand, the propagators in the first term enforce $\vb q_1 + \vb q_2 = \vb q_3 + \vb q_{4} = 0$, so the $\delta$-functions in \eqref{eq:form_of_S_I^2_terms} reduce to
\begin{equation}
  \delta^d(\vb k_1 + \vb k_{2}) \delta^d(\vb k_3 + \vb k_4).
\end{equation}
Transforming back to real space, the two $\delta$-functions guarantee that these contributions result in non-local terms of the form ${[ \int \dd[d]{x} (\grad\hat{\phi}^-\vdot\grad\phi^-)_{\vb x} ]^2}$. Such terms are cancelled by the corresponding terms in $\langle S_I \rangle_+^2$. 

There remains only the third term on the right-hand side of \eqref{eq44}. Inserting the relevant propagators into \eqref{eq:form_of_S_I^2_terms} and integrating over the $\delta$-function that enforces $\vb q_2 = -(\vb k_1 + \vb k_2 + \vb q_1)$ gives
\begin{equation}
  \delta^d(\vb k_1 + \vb k_2 + \vb k_3 + \vb k_4).
\end{equation}
After relabeling $\vb q_1 \rightarrow \vb q$, the second-order contribution to the effective action is found to be
\begin{align}
  \label{eq:second_order_contribution}
  &\frac{1}{2} (2^2)^2 g^2 \int_0^{\Lambda/\zeta} \left[\prod_{i=1}^4\frac{\dd[d]{k_i}}{(2 \pi)^d}\right] (\vb k_1 \cdot \vb k_2) (\vb k_3 \cdot \vb k_4) \nonumber\\
 &\times \hat{\phi}^-_{\vb k_1} \phi^-_{\vb k_2} \hat{\phi}^-_{\vb k_3} \phi^-_{\vb k_4} f(\vb k_1 + \vb k_2) (2\pi)^d \delta^d \left(\sum_{i=1}^4 \vb k_i \right),
\end{align}
where $f$ can be expanded as
\begin{align} \label{eq:expanding_f_order_k}
 & f(\vb k_1 + \vb k_2) = \int_{\Lambda/\zeta}^{\Lambda} \frac{\dd[d]{q}}{(2 \pi)^d} (\vb q \cdot (\vb k_1 + \vb k_2) + q^2)^2 \nonumber \\
  & \quad \times \frac{1}{q^2 \Kbar} \frac{1}{\abs{\vb k_1 + \vb k_2 + \vb q}^2 \Kbar}\nonumber\\
  &\qquad= \int_{\Lambda/\zeta}^{\Lambda} \frac{\dd[d]{q}}{(2 \pi)^d} (\Kbar)^{-2} [ 1 + O(\vb k_1 + \vb k_2) ].
\end{align}
The terms of $O(\vb k_1, \vb k_2)$ will generate terms in the effective action with two extra derivatives, 
\begin{equation}
k^6 (\hat{\phi}^-)^2(\phi^-)^2 \sim (\nabla \hat{\phi}^- \cdot \nabla \phi^-) \nabla^2 (\nabla\hat{\phi}^-\cdot\nabla\phi^-)\, ,
\end{equation} 
which are not of interest to us.

Including second-order contributions, the effective action is
\begin{widetext}
\begin{align}\label{eq3.53}
  &S'[\phi^-, \hat{\phi}^-] \approx -\int_0^{\Lambda / \zeta} \frac{\dd[d]{k}}{(2\pi)^d} k^2 \left[\Kbar - 4 \biggl(\frac{g}{\Kbar}\biggr) \int_{\Lambda/\zeta}^\Lambda \frac{\dd[d]{q}}{(2\pi)^d} \right] \hat{\phi}^-_{-\vb k} \phi^-_{\vb k} \nonumber\\
 & + \int_0^{\Lambda/\zeta} \left[\prod_{i=1}^4\frac{\dd[d]{k_i}}{(2 \pi)^d}\right] \left[g - \frac{1}{2} (2^2)^2 \biggl(\frac{g}{\Kbar}\biggr)^2 \int_{\Lambda/\zeta}^\Lambda \frac{\dd[d]{q}}{(2 \pi)^d}\right] 
   (\vb k_1 \cdot \vb k_2) (\vb k_3 \cdot \vb k_4) \hat{\phi}^-_{\vb k_1} \phi^-_{\vb k_2} \hat{\phi}^-_{\vb k_3} \phi^-_{\vb k_4} (2\pi)^d \delta^d \left(\sum_{i=1}^{4} \vb k_i\right)\, .
\end{align}
\end{widetext}
By comparing this result with the initial action, the effective $\Kbar'$ is unchanged from $O(g)$ in (\ref{eq2.42}). However, we obtain a first correction to $g$:
\begin{equation}
  g' = g - 8 \biggl(\frac{g}{\Kbar}\biggr)^2 \int_{\Lambda/\zeta}^{\Lambda} \frac{\dd[d]{q}}{(2\pi)^d}.
\end{equation}

\subsection{Rescaling and RG equations}%
\label{sub:rescaling_and_rg_equations}

To complete the RG transformation, the upper limit of integration must be restored to $\Lambda$ by rescaling the wave vector $\vb k' = \vb k \zeta$, which corresponds to a spatial rescaling of $\vb x' = \vb x/\zeta$.
The field rescaling of the RG transformation is typically carried out to maintain the functional form of the quadratic term $S_0$ \cite{tauber14}.
Suppose that we redefine the field with some power of the rescaling factor:
\begin{equation}
  \phi'_{\vb k} = \zeta^{-\omega_1} \phi^-_{\vb k/\zeta}, \qquad \hat\phi'_{\vb k} = \zeta^{-\omega_2} \hat\phi^-_{\vb k/\zeta}.
\end{equation}
We may also allow the mean permittivity $\Kbar$ to obey a scaling relation:
\begin{equation}
\Kbar^\prime= \zeta^{-\alpha} \Kbar.
\label{eq3.56}
\end{equation}
If $\alpha \neq 0$, then the effective mean permittivity is never finite in the limit of global upscaling $\zeta \rightarrow \infty$. In most applications and models where such an effective permittivity is desired, one will thus work with $\alpha = 0$. Nevertheless, we include the possibility of a nontrivial exponent $\alpha$ for purposes of generality. The exponents $\omega_1$ and $\omega_2$ can then be found by restoring the original integral limits of the quadratic term in the action through a change of variable $\vb k' = \vb k \zeta$:
\begin{align}
  &S_0[\phi^-, \hat\phi^-] = -\int_0^{\Lambda/\zeta} \frac{\dd[d]{k}}{(2\pi)^d} k^2 \Kbar \hat\phi^-_{-\vb k}\phi^-_{\vb k} \nonumber\\
  &\quad= -\int_0^\Lambda \frac{\dd[d]{k'}}{(2\pi \zeta)^d} \frac{k'^2}{\zeta^2} \zeta^\alpha \Kbar \zeta^{\omega_1 + \omega_2} \hat\phi'_{-\vb k'} \phi'_{\vb k}.
\end{align}
Requiring this to be of the original form, yields
\begin{equation}
  \label{eq3.58}
  \omega_1 + \omega_2 + \alpha = d + 2.
\end{equation}
Under the same change of variables, the interaction term rescales as
  \begin{align} \label{eq:S_I_rescaling_delta_cor}
     & S_I[\phi^-, \hat\phi^-] \nonumber\\
     &= \int_0^{\Lambda/\zeta} \left[\prod_{i=1}^4 \frac{\dd[d]{k_i}}{(2 \pi)^d}\right] g (\vb k_1 \cdot \vb k_2)(\vb k_3 \cdot \vb k_4) \nonumber\\
     & \quad {} \times \hat\phi^-_{\vb k_1} \phi^-_{\vb k_2} \hat\phi^-_{\vb k_3} \phi^-_{\vb k_4}(2\pi)^d \delta^d \left(\sum_{i=1}^4 \vb k_i \right)\nonumber\\
     &= \int_0^{\Lambda} \left[\prod_{i=1}^4 \frac{\dd[d]{k'_i}}{(2 \pi \zeta)^d}\right] g \frac{(\vb k'_1 \cdot \vb k'_2)(\vb k'_3 \cdot \vb k'_4)}{\zeta^4} \nonumber\\
      & \quad {} \times \zeta^{2(\omega_1 + \omega_2)} \hat\phi'_{\vb k'_1} \phi'_{\vb k'_2} \hat\phi'_{\vb k'_3} \phi'_{\vb k'_4}(2\pi \zeta)^d \delta^d\left(\sum_{i=1}^4 \vb k'_i\right),
  \end{align}
where the scaling factors incurred by the $d$-dimensional $\delta$-function were obtained from
  \begin{equation}
    \delta^d(\zeta^{-1}\vb k')   = \zeta^d \delta^d(\vb k').
  \end{equation}
Hence, by using (\ref{eq3.58}), the coupling constant is forced to rescale as
\begin{equation}
  g' = g \zeta^{[-4d-4+2(\omega_1 + \omega_2) + d]} = g \zeta^{-d - 2\alpha}.
\end{equation}
Thus, taking into account this scaling factor and that in (\ref{eq3.56}), the RG equations obtained by coarse-graining corrections up to $O(g^2)$ in the effective action are
\begin{align}
  \label{eq:RG1}
  \Kbar(\zeta) &= \zeta^{-\alpha} \left[\Kbar_0-4 \biggl(\frac{g_0 }{\Kbar_0}\biggr) \int_{\Lambda/\zeta}^{\Lambda} \frac{\dd[d]{q}}{(2\pi)^d} \right]\, , \\
  \label{eq:RG2}
  g(\zeta) &= \zeta^{-(d + 2\alpha)} \left[ g_0 - 8\biggl(\frac{g_0 }{\Kbar_0}\biggr)^2 \int_{\Lambda/\zeta}^{\Lambda} \frac{\dd[d]{q}}{(2\pi)^d} \right],
\end{align}
where the notation $\Kbar\equiv \Kbar(\zeta)$ and $g\equiv g(\zeta)$ signifies that the renormalized coupling constants depend on the scale of the coarse-graining as captured by $\zeta$. Their initial values are $\Kbar (1) \equiv K_0$ and $g (1) \equiv g_0$.

\subsection{DRG equations}%
\label{sub:drg_equations}

The DRG equations, often called $\beta$-functions 
, encapsulate what effect integrating out infinitesimal shells in momentum space has on the coupling constants $K$ and $g$.
We derive the $\beta$-functions from the RG equations \eqref{eq:RG1} and $\eqref{eq:RG2}$ by changing the scale variable to $\zeta = e^{s}$ and taking the limit $s \rightarrow 0$ at the end of our calculations.

The integral over $\dd[d]{q}$ appearing in both RG equations can then be expanded to first order in $s$ as an integral over the $d$-sphere in momentum space:
\begin{align}
  &\int_{\Lambda / \zeta}^\Lambda \frac{\dd[d]{q}}{(2 \pi)^d} = \int_{\Lambda e^{-s}}^{\Lambda} \frac{S_{d-1}}{(2\pi)^d} q^{d-1} \dd{q} \nonumber \\
  &\approx \frac{S_{d-1}}{(2\pi)^d}( \Lambda - \Lambda e^{-s} )\Lambda^{d-1}
  \approx S_{d-1} \left( \frac{\Lambda}{2\pi} \right)^d s,
  \label{eq64}
\end{align}
where $S_d$ is the surface area of the unit $d$-sphere. This expansion becomes exact in the limit $s \rightarrow 0$. Defining
\begin{equation}
C_d \equiv S_{d-1}\biggl(\frac{\Lambda}{2\pi}\biggr)^d.
\end{equation}
and substituting (\ref{eq64}) into the RG equations gives
\begin{align}
  \Kbar(s) &\approx e^{-\alpha s} \left[\Kbar_0- 4C_d \biggl(\frac{g_0 }{\Kbar_0}\biggr) s \right]\, , \\
  g(s) &\approx e^{-(d + 2\alpha ) s} \left[ g_0 - 8C_d \biggl(\frac{g_0 }{\Kbar_0}\biggr)^2 s\right].
\end{align}
The RG transformations restore the form of the action $S$ by replacing the coupling constants $\Kbar_0$ and $g_0$ with effective, scale-dependent values $\Kbar (s)$ and $g(s)$. Hence, a second application of the RG, coarse-graining from $s$ up to $s + \delta s$, gives RG equations of the same form as above:
\begin{align}
  \Kbar(s + \delta s) &\approx e^{-\alpha \delta s} \left[\Kbar (s) - 4C_d \biggl(\frac{g (s)}{\Kbar (s)}\biggr) \delta s \right]\, , \\
  g(s + \delta s) &\approx e^{-(d + 2\alpha ) \delta s} \left[ g (s) - 8C_d \biggl(\frac{g (s)}{\Kbar (s)}\biggr)^2 \delta s \right].
\end{align}
In the limit $\delta s\to 0$, we obtain the DRG equations
\begin{align}
  \label{eq:DRG1}
  \dv{\Kbar}{s} &= -\alpha \Kbar - 4C_d \left( \frac{g}{\Kbar} \right) \\
  \label{eq:DRG2}
  \dv{g}{s} &= -(d + 2\alpha) g - 8C_d \left( \frac{g}{\Kbar} \right)^2.
\end{align}
The fixed points and exact trajectories of these equations will be obtained in the next section.

\section{RG Trajectories}%
\label{sec:rg_trajectories_and_fixed_points}

\subsection{Fixed Points}%
\label{sub:rg_flow_and_fixed_points}

The fixed points of the system, where $\Kbar^\prime=0$ and $g^\prime=0$, are obtained by simultaneously solving
\begin{align}
   &\alpha \Kbar + 4C_d \frac{g}{\Kbar}=0 \, , \\
  (d + 2&\alpha)g + 8C_d \left( \frac{g}{\Kbar} \right)^2=0 \, .
\end{align}
For $\alpha = 0$, there is a fixed point at $g^\ast = 0$ for all non-zero values of $\Kbar$. As we discussed previously, $\alpha = 0$ is the only value of the scaling exponent for which we obtain a finite effective conductivity in the asymptotic limit. In addition, the scaling analysis and the $\beta$-function for $g$ show that the RG flow will, for any dimension $d$, carry the coupling constants to this fixed point:~$g(s \rightarrow \infty)\equiv g^\ast = 0$. The coupling constant $g$ was defined in (\ref{eq24}) to be proportional to the variance $\sigma^2$. Thus, this fixed point, reached for all $d$ in the limit $s \rightarrow \infty$, corresponds to the limit $\sigma^2 \rightarrow 0$ of small conductivity fluctuations. 

The linearized RG equation for $g$ about the fixed point is
\begin{equation}
  \dv{g}{s} = -dg ,
\end{equation}
yields, with $G(0)=g_0$,
\begin{equation}
  g(s)=g_0e^{-ds}.
\end{equation}
By inserting this into the differential equation for $\Kbar$, we obtain
\begin{equation}
  \dv{\Kbar}{s} = -4C_d \frac{g_0}{\Kbar} e^{-ds}.
\end{equation}
By employing separation of variables to solve this ODE and using the initial condition $\Kbar (0) = \Kbar_0$, the effective permittivity as a function of the scale factor $\zeta = e^{s}$ is:
\begin{equation}
  \Kbar(s) = \sqrt{ \Kbar_0^2 - \frac{8C_d g_0}{d}(1 - e^{-ds})}.
\end{equation}
Reinstating the constant $C_d$, and recalling the definition (\ref{eq24}) of the coupling constant, the effective conductivity and variance for the coarse grained theory at length scale $\vb x (\zeta) = \vb x_0 / \zeta$ are
\begin{align}
  \Kbar(\zeta) &= \sqrt{\Kbar_0^2 - \frac{4 S_{d-1}}{d (2\pi)^{d/2}} \bigl[\sigma_0^2 - \sigma^2(\zeta)\bigr]}\, , \\
  g^2(\zeta) &= g_0 \zeta^{-d}\, ,\qquad \sigma^2(\zeta) = \sigma_0^2 \zeta^{-d}\, .
\end{align}
These solutions of the linearized DRG equations are valid close to the fixed point, where $\sigma^2$ is small. At the fixed point,
\begin{equation}
  \Kbar^\ast = \sqrt{\Kbar_0^2 - \frac{4 S_{d-1}}{d (2\pi)^{d/2}} \sigma_0^2}.
\end{equation}
In $d=2$, with $S_1 = 2\pi$,
\begin{align}
  \Kbar_{d=2}^\ast= \sqrt{\Kbar_0^2 - 2 \sigma_0^2}
  \approx \Kbar_0 \left[ 1 - \left( \frac{\sigma_0}{\Kbar_0} \right)^2 \right].
\end{align}

\subsection{Exact Solutions}%
\label{sub:exact_solutions}

The exact solutions to the DRG equations \eqref{eq:DRG1} and \eqref{eq:DRG2} are:
\begin{align}
  \Kbar(s) &= C_1 \exp(C_2 e^{-ds} - \alpha s) \label{eq:Kbar_exact} \\
  g(s) &= \frac{C_2 d}{4 C_d} \Kbar^2 e^{-ds} \label{eq:g_exact} \, ,
\end{align}
where $C_{1, 2}$ are constants of integration that are to be determined by initial conditions.
A derivation of these solutions is given in Appendix \ref{apx:exact_solutions}. Enforcing the boundary conditions $\Kbar(0) = \Kbar_0$ and $g(0) = g_0$, these solutions, written in terms of the rescaling factor $\zeta = e^s$, are
\begin{align}
    \Kbar(\zeta) &= \zeta^{-\alpha} \Kbar_0 \exp{\frac{4C_d}{d} \left( \frac{g(s) - g(0)}{\Kbar_0^2} \right)} \\
    g(\zeta) &= g_0 \zeta^{-d}\, , \qquad  \sigma^2(\zeta) = \sigma^2_0 \zeta^{-d} .
\end{align}
A finite mean conductivity in the coarse-grained limit $\zeta \to \infty$ requires choosing $\alpha = 0$.

\section{Finite Correlation Length}
\label{sec5}

Our analysis in the preceding sections has been based on Gaussian conductivity fluctuations with a delta correlation function. This provides a basic case for determining the structure of RG transformations for Darcy's law with a stochastic conductivity, but a more realistic description of conductivity fluctuations must be examined to apply our results to physical systems.

Correlations have a pronounced effect of the variations of the conductivity \cite{westbroek19a}, so in this section, we introduce a finite correlation length to the fluctuations to determine its effect on the RG transformations. We consider a system with a finite isotropic correlation length $l$ by defining the correlation function as
\begin{equation}
  C(\vb{x}, \vb{y}) = \langle \widetilde K(\vb{x}) \widetilde K(\vb{y}) \rangle= \sigma^2 \exp\biggl({-\frac{\abs{\vb{x} - \vb{y}}^2}{2 l^2}}\biggr)\, .
  \label{eq87}
\end{equation}
Inserting this correlation function into \eqref{eq:action_as_fn_of_C} gives the new action
\begin{multline}
  S[\phi, \hat\phi] = -\int \dd[d]{x} \Kbar \nabla \hat\phi_{\vb x} \cdot \nabla\phi_{\vb x} \\
  {}+\frac{\sigma^2}{2} \int \dd[d]{x} \dd[d]{y} (\nabla \hat\phi \cdot \nabla\phi)_{\vb x} (\nabla \hat\phi\cdot \nabla \phi)_{\vb y} e^{-\frac{\abs{\vb x - \vb y}^2}{2l^2}}\, .
\end{multline}
The quadratic part $S_0$ of the action is the same as that in \eqref{eq:S_I_fourier_zero_corlen} because only the mean of the conductivity enters this term. The interaction part $S_I$ of the action, which embodies conductivity fluctuations, is augmented by the exponential factor containing the correlation length in (\ref{eq87}). The Fourier representation of the new interaction term is
\begin{widetext}
\begin{align} \label{eq:S_I_fourier_space_finite_cor_len}
  S_I[\phi, \hat\phi] &= \frac{\sigma^2}{2}  \int \qty[\prod_{i=1}^4 \frac{\dd[d]{k_i}}{(2 \pi)^d}] (\vb k_1 \cdot \vb k_2)(\vb k_3 \cdot \vb k_4) \hat\phi_{\vb k_1}\phi_{\vb k_2}\hat\phi_{\vb k_3}\phi_{\vb k_4} (2 \pi)^d \delta^d \left(\sum_{i=1}^4 \vb k_i \right)\nonumber \\
  &\quad\times (2 \pi l_0^2)^{d/2} \exp[ \frac{(\vb k_1 + \vb k_2)\cdot(\vb k_3 + \vb k_4)}{2 l^{-2}} ].
\end{align}
\end{widetext}
To compare this expression with our previous action, we define the coupling constant to include the correlation length $l$ instead of the smallest length scale $\Lambda^{-1}$ in the system:
\begin{equation}
  g_l = \frac{1}{2} \sigma^2  (2\pi l^2)^{d/2}.
\end{equation}
Using this definition, and comparing \eqref{eq:S_I_fourier_space_finite_cor_len} and \eqref{eq:S_I_fourier_zero_corlen}, we see that the finite correlation length $l$ brings an additional exponential factor
\begin{equation}
  \label{eq:exponential_factor}
  \exp[\frac{(\vb{k}_1 + \vb{k}_2) \cdot (\vb{k}_3 + \vb{k}_4)}{2l^{-2}}].
\end{equation}
Due to the momentum conserving $\delta$-function in the action, the wave vectors in the exponential factor can be  replaced by
\begin{equation}
  \vb{k}_1 + \vb{k}_2 = -(\vb{k}_3 + \vb{k}_4).
\end{equation}
However, we chose the representation above to manifestly show the symmetry between the two $\vb k_i \cdot \vb k_j$ pairs. In the limit of $l \to \infty$, the exponential factor becomes a second $\delta$-function
\begin{align}
  &\lim_{l^{-1} \rightarrow 0} \left[ (2\pi)^d (2\pi l^{-2})^{-d/2} \exp(-\frac{\abs{\vb k_3 + \vb k_4}^2}{2 l^{-2}} ) \right] \nonumber \\
  &\qquad = (2\pi)^d \delta^d(\vb k_3 + \vb k_4),
\end{align}
where, as noted above, the argument of the $\delta$-function can also be chosen as $\vb{k}_1 + \vb{k}_2$.
In the limit $l \rightarrow \Lambda^{-1}$, the exponential factor becomes
\begin{multline} \label{eq:vanishing_cor_len}
(2\pi \Lambda^{-2})^{d/2} \exp(-\frac{\abs{\vb k_3 + \vb k_4}^2}{2\Lambda^{2}})\\
\approx (2\pi)^{d/2} \Lambda^{-d} + O(\Lambda^{-(d + 2)}).
\end{multline}
In this limit, $g_l \to g$, and we recover the action \eqref{eq:S_I_fourier_zero_corlen} for the case of vanishing correlation length.
In fact, this is how we chose the normalization for the $\delta$-function correlation $C(\vb{x}, \vb{y})$ in \eqref{eq:delta_correlation}, observing that it arises naturally from the limit $l \to \Lambda^{-1} \to 0$ of vanishing correlation length:
\begin{align}
  &C(\vb{x}, \vb{y}) = \sigma^2 \exp\biggl(-\frac{\abs{\vb{x} - \vb{y}}^2}{2 l^2}\biggr) \nonumber \\
  &\xrightarrow{l \to \Lambda^{-1}} \sigma^2 \left[ \frac{2\pi \Lambda^{-2}}{2\pi \Lambda^{-2}}^{d/2} \exp\biggl(-\frac{\abs{\vb{x} - \vb{y}}^2}{2 \Lambda^{-2}}\biggr) \right] \nonumber\\
  &\xrightarrow{\Lambda^{-1} \to 0} \sigma^2 (2\pi \Lambda^{-2})^{d/2} \delta^d (\vb{x} - \vb{y}).
\end{align}

\subsection{Effective Action}%
\label{sub:effective_action_finite_corlen}

\subsubsection{First Order}%
\label{subsub:first_order_finite_corlen}

To account for the extra exponential factor resulting from the finite correlation length, we invoke the derivations in Sec.~\ref{sub:integrating_out_high_wave_numbers}. In particular, the first-order contribution $\langle S_I \rangle_+$ to the effective action is the same as \eqref{eq3.40}, except for the additional exponential factor in the integral. Moreover, when we perform the integral over $\dd[d]{k_3}$, using the $\delta$-function to replace $\vb k_3 = -\vb k_4$, this exponential factor vanishes again. Therefore, the resulting first order contribution is the same as that obtained for vanishing correlation length, except that now $\Lambda^{-1} \to l$.

\subsubsection{Second Order}%
\label{subsub:second_order_finite_corlen}

The second-order contribution, which has the general form (\ref{eq44a}), can be similarly obtained by multiplying the integrand of \eqref{eq:form_of_S_I^2_terms} by two exponential factors of the form in \eqref{eq:exponential_factor}, one each for the integrals over the $k_i$ and the $q_i$. Proceeding in the same way as the previous derivation using Wick's theorem, we find that the second-order contribution with finite correlation length is the same as \eqref{eq:second_order_contribution}, except for an additional exponential factor, which can again be expanded as $1 + O\bigl[(\vb k_1 + \vb k_2)^2\bigr]$. Using the same argument as before, the contributions of these $O(k^2)$ terms to the effective action are of no interest to us. Therefore, to $O(k^2)$, the additional exponential incurred by the finite correlation length does not influence the RG equations for $\Kbar$ and $g$ to first or second order. However, the finite correlation length does change the scaling of the variance.

\subsubsection{Rescaling}%
\label{subsub:rescaling_finite_corlen}

Because a finite correlation length does not alter the quadratic term of the action, we use the same relation \eqref{eq3.58} to rescale the fields. Let us assume that $\alpha = 0$ for this section, meaning that the average conductivity $\Kbar$ is assumed to be scale independent.
Rescaling the coordinates $\vb{k} \to \vb{k}' = \vb{k}\zeta$ to restore the UV cutoff, the interaction term rescales as in \eqref{eq:S_I_rescaling_delta_cor}, necessitating a rescaling of the coupling constant as $g\to g' = \zeta^{-d} g$. However, we must now also consider the rescaling of the new exponential factor
\begin{align}
  &\exp[\frac{(\vb{k}_1 + \vb{k}_2) \cdot (\vb{k}_3 + \vb{k}_4)}{2 l^{-2}}]\nonumber\\ 
  &\quad=\exp[\frac{(\vb{k'}_1 + \vb{k'}_2) \cdot (\vb{k'}_3 + \vb{k'}_4)}{2 \zeta^2 l^{-2}}]\, .
\end{align}
This suggests that we should rescale the correlation length as $l \to l' = l / \zeta$. However, since the coupling constant $g_l = \frac{1}{2} \sigma^2 (2\pi l^2)^{d/2}$ contains a factor of $l^d$, this correlation length rescaling would take care of the entire rescaling $g' = \zeta^{-d} g$. In particular, the variance $\sigma^2$ would not need to be rescaled; in this theory the correlation length shrinks under coarse-graining while the variance is scale invariant! This is very different from the limit $l \to \Lambda^{-1}$ of vanishing correlation length, where the rescaling of $g$ was taken to be equivalent to rescaling $\sigma^2$.

\subsubsection{Effective Conductivity}%
\label{subsub:effective_conductivity}

In the case of finite correlation length, the RG equations are the same as the ones for delta function correlation.
The only difference lies in the interpretation of the rescaling of $g_l$, as discussed in the preceding subsection.
In particular, the effective conductivity in the vicinity of the fixed point $g_l=0$ is
\begin{equation}
  \Kbar(\zeta) = \sqrt{ \Kbar_0^2 - \frac{4S_{d-1} \sigma_0^2 (l_0 \Lambda)^d)}{d (2\pi)^{d/2}} \bigl(1-\zeta^{-d})}.
\end{equation}
This effective action, which has been obtained by using the RG transformations to integrate out wavenumbers between $\Lambda / \zeta < \Lambda$ and the UV cutoff $\Lambda$, describes a coarse-grained theory of Darcy flow. This coarse-grained theory accurately describes processes on large scales, whenever small-scale fluctuations (of length-scale $\sim \zeta/\Lambda$) cannot be resolved.

\section{Summary and Future Work}
\label{sec6}

We have used the path integral formulation for the solution for Darcy's equation with a stochastic conductivity for the flow of a single fluid through a random porous medium.  Beginning with the case of delta-correlated Gaussian fluctuations, we obtained differential RG equations to second order in the coupling constant.  These equations were used to identify the fixed points, physically appropriate values of he scaling exponents, and exact solutions for the RG trajectories.  The extension to Gaussian fluctuations with a finite correlation length produced the same RG equations as the delta-correlated fluctuations;~the only differences were in the interpretation of the rescaling of the coupling constant.

There are several extensions of our study that are needed to make RG calculations a practical method for modelling flow through random porous media.  Comparisons with direct numerical evaluations of the path integral for Darcy flow will enable us to ascertain the accuracy of the expansions used in our calculations.  Work in this direction for Darcy's law is described in Refs.~\cite{attinger03,hanasoge17} and for classical diffusion in random media in Ref.~\cite{deem94}.  We have at our disposal calculations in 1, 2, and 3 dimensions \cite{westbroek19a,westbroek19b}, so we can investigate the dimensional dependence of the accuracy of our calculations.  As noted in the Introduction, RG equations in different spatial dimension have the same diagrammatic structure, but the methods used for the numerical valuation of the path integral are different in $d=1$ \cite{westbroek19a} and $d=2$ and $d=3$ \cite{westbroek19b}. This should provide an additional point of interest for our comparisons.

The calculations reported here are based on Gaussian fluctuations.  In fact, in geological applications, the logarithm of the conductivity is assumed to have a Gaussian distribution \cite{law44,li05}.  In principle, any probability distribution function of the conductivity can be described in terms of the characteristic function \cite{phythian77},  but there will be substantial changes to the RG equations.  On the other hands, we expect that for length scales longer than the correlation length of fluctuations, whatever their distribution, under the coarse-graining transformation, the distribution will approach a Gaussian.  The challenge is to determine the intermediate distributions.

The final extension, to multiphase flow, is a more substantial endeavor.  Multiphase flow through porous media is important for a various applications, such as CO$_2$ sequestration \cite{juanes12}, and enhanced oil recovery \cite{orr84}. These often involve the displacement of a non-wetting invading fluid from a porous medium by a wetting fluid (imbibition).  For two-phase flow a generalized form of Darcy's law is used:
\begin{equation}\label{Darcy_multiphase}
{\bf q}_i = -k_{r,i}(S_i)K\nabla p\, ,
\end{equation}
where the subscript $i$ represents the phases of the fluids, $k_{r,i}$ is the relative permeability, and $S_i$ is the pore volume fraction of the fluid phase $i$.  The two volume fractions must sum to one. The total velocity is given by the sum of the individual phase velocities:
\begin{equation}
{\bf q} = {\bf q}_o + {\bf q}_w\, .
\end{equation}
The rate of change of the saturation $s$ is given by the conservation equation:
\begin{equation}\label{conservation}
\frac{\partial s}{\partial t}=g(s) {\bf q} \cdot \nabla s\, ,
\end{equation}
where $g(s)$ is a nonlinear function. The constraint $\nabla\cdot{\bf q}=0$ still holds for an incompressible fluid. Equation (\ref{Darcy_multiphase}) is similar to Darcy's law for single-phase flow.  An adaptation of the methodology outlined in this paper should be suited to the solution of (\ref{Darcy_multiphase}).  The hyperbolic saturation equation (\ref{conservation}) poses more problems, in particular because its nonlinear nature leads to the formation of a shock in the solution \cite{buckley42}. Previous studies \cite{king02} indicate how a path integral solution of this equation can be formed, providing the first step of an RG analysis.

\begin{acknowledgments}
U.\"O.~acknowledges the support by a bursary from the Undergraduate Research Opportunities Programme (UROP) at Imperial during the summer of 2019.
\end{acknowledgments}

\pagebreak[1]
\appendix
\section{Feynman Diagrams}
\label{apx:diagrams}

\subsection{Rules}%
\label{sub:rules}

The perturbative calculations of the first and second contributions to the effective action can be considerably simplified by the use of Feynman diagrams.  This diagrammatic technique, which helps keep track of the different integrals, is especially useful when calculating higher order contributions for which Wick's theorem becomes increasingly cumbersome.

The following rules have been adapted from the usual Feynman diagram rules to account for the derivative terms in our action $S_I$. 
Each diagram corresponds to an integral contributing to the perturbative expansion of the Wilsonian effective action of \eqref{eq:effective-action}. The fact that we are dealing with an expansion of $\ln \langle e^{-S_I} \rangle$ means that we will only have to consider connected diagrams. 

\begin{figure*}[tbph]
\centering
\includegraphics[width=0.8\textwidth]{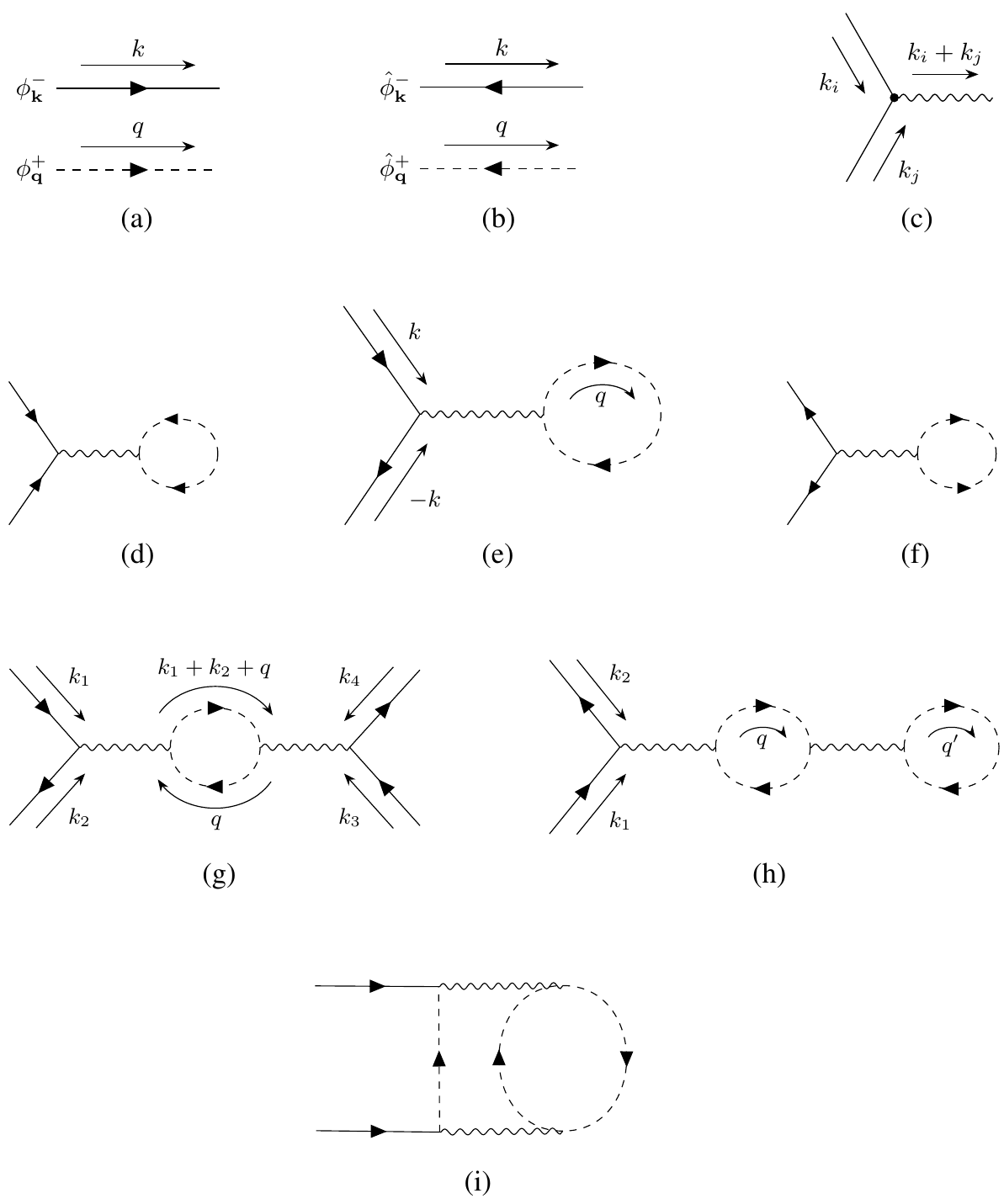}
\caption{Diagrams that illustrate the rules up to second order in the coupling constant for the contributions to the effective action.  (a,b) Rules for lines and (c) vertices.  (d,e,f) First-order diagrams, and (g,h,i) second-order diagrams. Solid lines correspond to long-wavenumber fields and dotted lines to low wavenumber fields}
\label{fig1}
\end{figure*}

To construct a diagram, one begins by choosing an arbitrary direction to be the principal direction of the diagram; in all following diagrams, this direction will be from left to right, so an explicit specification of this `time'-direction is omitted. One can then draw solid lines, which correspond to the low wavenumber fields $\phi^-_{\vb{k}}$ and $\hat \phi^-_{\vb{k}}$, and dotted lines, which denote the high wavenumber fields $\phi^+_{\vb{q}}$ and $\hat \phi^+_{\vb{q}}$ (Fig.~\ref{fig1}(a,b)). The fact that we are integrating over the high wavenumber modes is reflected diagrammatically in the constraint that all dotted lines have to be internal, meaning that both ends of the dotted lines have to be connected to a vertex.  Moreover, each internal line with momentum $\vb{q}$ contributes a propagator factor $G(q)$ and each closed loop is integrated over, contributing an integral $\int_{}^{} \dd[d]{q} (2\pi)^{-d}$. In contrast, solid lines are allowed to be external, meaning that they are only connected to a single vertex. This reflects the fact that the solid lines correspond to the low wavenumber modes, which remain in the theory after the RG coarse-graining.
For both the internal and external fields, the lines that represent conjugate fields $\hat \phi^{\pm}$ are to be distinguished from their counterparts $\phi^{\pm}$ by being drawn with reversed arrows, pointing opposite the principal direction of the diagram. 

The momenta, denoted $\vb{k}$ for external and $\vb{q}$ for internal fields, are specified on a separate arrow, as illustrated in Fig.~\ref{fig1}(a,b).
The vertices group the momenta into pairs $\vb{k}_i \cdot \vb{k}_j$ as depicted in Fig.~\ref{fig1}(c). Each such vertex contributes a factor of ${\sqrt{g} (\vb{k}_i \cdot \vb{k}_j)}$ to the integral. This factor was chosen so that a complete interaction between four fields $S_I \sim (\grad \phi^-_{\vb{k}} \cdot \grad \phi^-_{\vb{k}})^2$ contributes a factor of ${g (\vb{k}_1 \cdot \vb{k}_2) (\vb{k}_3 \cdot \vb{k}_4)}$. The wavy line, which separates out each four-field interaction into a pair of two-point vertices, is introduced solely to give us the ability to group the four momenta of each interaction into two pairs; unlike in quantum field theory, where the interactions are mediated via gauge bosons, we assign no physical significance to these wavy lines. The momenta flowing into each vertex are required to add up to zero. Overall, the integral corresponding to each diagram will have a momentum conserving delta function factor $(2\pi)^d \delta^d(\sum_i \vb{k}_i)$, where the sum is over all the external fields $\phi^-_{\vb{k}_i}$.

Finally, to account for permutations of lines that yield identical diagrams, we multiply the contribution of each diagram by its symmetry factor, which is the order of its automorphism group; in other words, the symmetry factor is obtained by counting the number of diagrams which produce identical contributions. This factor counts the number of equivalent pairings which arise in Wick's theorem.

In the following two subsections, we will go through the one-loop diagrams with one and then two vertices, which correspond to contributions of order $O(g)$ and $O(g^2)$ respectively. We will highlight the diagrams of integrals we have already calculated using Wick's theorem, and will expand on these by discussing additional two-loop diagram terms that we have previously neglected. Although the formal rules outlined above should be used when performing calculations, we will drop the momentum labels when they are not essential to our understanding of the contribution of each diagram.

\subsection{First Order Diagrams}%
\label{sub:first_order_diagrams}

Since each vertex contributes a factor of $g$, we can keep track of the perturbative expansion by counting the number of vertices. Following the discussion of our approach using Wick's theorem, we first consider the trivial diagrams.

Consider first the diagram that contains only external lines. 
\begin{figure}[htbp]
\centering
\includegraphics[width=6cm]{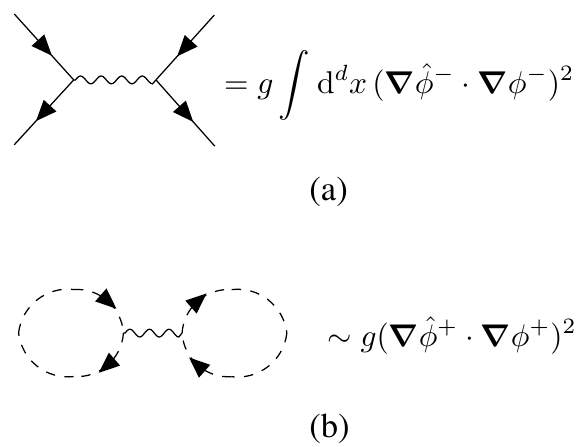}
\caption{(a) A diagram that contains only low wavenumber fields and is unaffected by coarse graining. This diagram contributes directly to the effective action. (b) A diagram with only internal lines, which contributes a constant to the effective action.}
\label{fig2}
\end{figure}
This diagram, which is shown in Fig.~\ref{fig2}(a), corresponds to the term that contains only low wavenumber fields and is, therefore, unaffected by the integral over high wavenumbers. The effect of coarse graining is simply to pass his term over to the effective action.

The second trivial diagram contains only internal lines, corresponding to high wavenumber fields (Fig.~\ref{fig2}(b)).  These get integrated over and contribute an irrelevant constant to the effective action. This exemplifies that only those diagrams that contain at least one pair of external solid lines contribute to the effective action.

The first non-trivial diagram that we can draw while respecting momentum conservation at each vertex has one loop. It is depicted in Fig.~\ref{fig1}(e).
Following our rules, this diagram evaluates to
\begin{equation}
  4 g \int_{\Lambda/ \zeta}^{\Lambda} \frac{\dd[d]{q}}{(2\pi)^d} q^2 G_0(q) \int_{}^{} \dd[d]{x} \grad \hat\phi^- \cdot \grad \hat \phi^-,
\end{equation}
where the symmetry factor $4$ arises because we can independently reflect the left legs or the right loop about the central axis of the diagram to obtain three other equivalent diagrams. This is equivalent to the statement that this diagram has $\mathbb{Z}_2 \times \mathbb{Z}_2$ symmetry.
Inserting the propagator $G_0(q)$, we see that this diagram is equivalent to \eqref{eq:order_g_wicks_theorem}, which was obtained algebraically by applying Wick's theorem. Note that the propagator can be non-zero only if every loop is made up of one $\phi^+$ and one $\phi^-$ field. 
Therefore, the only diagrams that we need to consider are those that have one line flowing in and one line flowing out of each vertex.
In particular, the contributions from diagrams shown in Figs.~\ref{fig1}(d) and \ref{fig1}(f) vanish, since they are proportional to $\langle \phi^+ \phi^+ \rangle$ and $\langle \hat\phi^+ \hat\phi^+ \rangle$, respectively.

\subsection{Second Order Diagrams}%
\label{sub:second_order_diagrams}

Since we are dealing with a cumulant expansion of $\ln \langle e^{-S_I} \rangle_+$, only connected graphs appear in the expansion. The $O(g^2)$ correction that we computed by means of Wick's theorem arises from the diagram shown in Fig.~\ref{fig1}(g). Using the Feynman diagram rules, this diagram corresponds to the integral
\begin{widetext}
\begin{align}
  &8g^2 \int_{0}^{\Lambda/\zeta} \left[ \prod_{i=1}^4 \frac{\dd[d]{k_i}}{(2\pi)^d} \right] 
  \Biggr\{ (\vb{k}_1 \cdot \vb{k}_2) (\vb{k}_3 \cdot \vb{k}_4) 
    \phi^-_{\vb{k}_1} \hat\phi^-_{\vb{k}_2}\phi^-_{\vb{k}_3} \hat\phi^-_{\vb{k}_4}
    (2\pi)^d \delta^d \left(\sum_{i=1}^4 \vb{k}_i\right)\nonumber \\
  &\quad\times \int_{\Lambda/\zeta}^{\Lambda} \frac{\dd[d]{q}}{(2\pi)^d} [\vb{q} \cdot (\vb{k}_1 + \vb{k}_2 + \vb{q})]^2 G_0(q) G_0\bigl(\abs{\vb{k}_1 + \vb{k}_2 + \vb{q}}\bigr) \Biggr\}.
\end{align}
\end{widetext}

However, at order $O(g^2)$, there are two further diagrams that we can build.
The first one is illustrated in Fig.~\ref{fig1}(h).  This diagram corresponds to the integral
\begin{align}
&8g^2 \int_{\Lambda/\zeta}^{\Lambda} \frac{\dd[d]{q}}{(2\pi)^d} q^2 G^2 (q) \int_{\Lambda/\zeta}^{\Lambda} \frac{\dd[d]{q'}}{(d\pi)^d} q'^2 G(q')\nonumber\\
&\quad\times \int \dd[d]{x} \grad \hat\phi^- \cdot \grad \phi^-\, ,
\end{align}
which shifts the quadratic term. Treating this term as an additional correction to the RG equation, the two integrals over $\dd[d]{q}$ and $\dd[d]{q'}$ give corrections proportional to $s^2$. Therefore, the DRG equations obtained from integration over infinitesimal $\delta s$ are unchanged under this two-loop diagram.

Similarly, the diagram in Fig.~\ref{fig1}(i) gives a contribution of the form
\begin{equation}
g^2 \int_{}^{} \frac{\dd[d]{k}}{(2\pi)^d} A(k, \Lambda) \phi^-_{\vb{k}} \hat\phi^-_{-\vb{k}}.
\end{equation}
This shifts both the quadratic term and the gradient term. This can be seen by expanding the function $A_k$ as a power series in $k$:
\begin{equation}
  A_k = A_0 + \frac{1}{2} k^2 A''_0 + \ldots.
\end{equation}
The two-loop diagrams change the propagator which then affect the couplings only at higher order. We therefore see that the DRG equations that we discussed here are the one-loop $\beta$-functions of the path integral renormalization of Darcy's law. To improve the accuracy of our predictions, an extension of this work could aim to compute the corrections arising from the one-loop diagrams with $n$-vertices, corresponding to $O(g^n)$ contributions.

\section{Exact Solutions of DRG Equations}%
\label{apx:exact_solutions}

We adopt a more convenient notation in which the DRG equations \eqref{eq:DRG1} and \eqref{eq:DRG2} are
\begin{align}
  \dot x_1 &= -a x_1 - b x_2 x_1^{-1}\, , \label{eq:I} \\
  \dot x_2 &= -c x_2 - 2b (x_2 x_1^{-1})^2 \, , \label{eq:II}
\end{align}
where $(x_1, x_2) = (\Kbar, g)$, $a  = \alpha$, $b = 4 C_d$, and $c = (d + 2 \alpha)$. The overdot denotes differentiation with respect to the renormalization parameter $t$.

We first multiply  \eqref{eq:I} by $x_1 \neq 0$, giving
\begin{equation}
  \dot{x}_1x_1 + ax_1^2 + b x_2 = 0.
\end{equation}
The term $\dot{x}_1x_1$ suggests the substitution $u = \frac{1}{2} x_1^2$, whereupon we have
\begin{equation}
  \dot{u} + 2 a u + b x_2 = 0.
\end{equation}
Differentiating with respect to $t$, we can use \eqref{eq:II} to substitute for $\dot{x}_2$, giving
\begin{align}
&\ddot{u} + 2 a \dot{u} + b \dot{x}_2 \nonumber\\
 &\quad= \ddot{u} + 2a \dot{u} - bc x_2 - 2b^2 x_2^2 x_1^{-2} = 0.
\end{align}
Using $x_1^2 = 2u$ and $bx_2 = -\dot{u} - 2 au$, we obtain a second-order differential equation for $u$:
\begin{equation}
  \label{eq:u-only}
  \ddot{u} + \dot{u}(-2 a + c) + u (2ac - 4 a^2) - \frac{\dot{u}^2}{u} = 0.
\end{equation}
Now the term  $\dot{u} / u$ suggests another substitution $y = \ln u$, whereupon \eqref{eq:u-only} becomes
\begin{equation}
  \ddot{y} + \dot{y} (c - 2a) + (2 a c - 4 a^2) = 0.
\end{equation}
Finally, we substitute $z = \dot{y}$ to obtain
\begin{equation}
  \dot{z} = (z + 2a) (2a - c),
\end{equation}
which can be solved by separation of variables:
\begin{equation}
  \ln \abs{z + 2 a} = (2 a- c) t + A_1,
\end{equation}
where $A_1$ is a constant of integration.
Back substituting $z = \dot{y}$, and absorbing the sign from the absolute value into the integration constant, we find $\dot{y} = A_1' e^{(2a - c)t} - 2a$. Integrating gives
\begin{equation}
  y = A_1'' e^{(2a - c)t} - 2 at + A_2,
\end{equation}
where $A_2$ is another integration constant.

After performing the final two back substitutions $u = e^y$, followed by  $x_1 = \pm \sqrt{2 u}$, we have
\begin{equation}
  \label{eq:x_1}
  x_1(t) = C_1 \exp[C_2 e^{(2a - c) t} - at],
\end{equation}
where all factors have been absorbed into the final integration constants $C_{1, 2}$.
Finally, we use \eqref{eq:I} to solve for $x_2$, giving
\begin{align}
  \label{eq:x_2}
  x_2(t) &= -\frac{C_1^2 C_2}{b} (2 a - c) e^{(2 a - c)t} \nonumber \\
	 &\qquad \times \exp[2 C_2 e^{(2a - c)t} - 2at] \nonumber \\
	 &= - \frac{x_1^2}{b} C_2 (2 a - c) e^{(2a - c) t} \, ,
\end{align}
The constants $C_{1, 2}$ can be fixed by enforcing initial conditions for $x_1, x_2$.
One may check by simple substitution that \eqref{eq:x_1} and \eqref{eq:x_2} indeed satisfy \eqref{eq:I} and \eqref{eq:II}.
Reinstating the constants $a$, $b$, and $c$, these are the equations \eqref{eq:Kbar_exact} and \eqref{eq:g_exact} that are discussed in Sec.~\ref{sub:exact_solutions}.

\bibliography{manuscript}

\begin{thebibliography}{49}%
\makeatletter
\providecommand \@ifxundefined [1]{%
 \@ifx{#1\undefined}
}%
\providecommand \@ifnum [1]{%
 \ifnum #1\expandafter \@firstoftwo
 \else \expandafter \@secondoftwo
 \fi
}%
\providecommand \@ifx [1]{%
 \ifx #1\expandafter \@firstoftwo
 \else \expandafter \@secondoftwo
 \fi
}%
\providecommand \natexlab [1]{#1}%
\providecommand \enquote  [1]{``#1''}%
\providecommand \bibnamefont  [1]{#1}%
\providecommand \bibfnamefont [1]{#1}%
\providecommand \citenamefont [1]{#1}%
\providecommand \href@noop [0]{\@secondoftwo}%
\providecommand \href [0]{\begingroup \@sanitize@url \@href}%
\providecommand \@href[1]{\@@startlink{#1}\@@href}%
\providecommand \@@href[1]{\endgroup#1\@@endlink}%
\providecommand \@sanitize@url [0]{\catcode `\\12\catcode `\$12\catcode
  `\&12\catcode `\#12\catcode `\^12\catcode `\_12\catcode `\%12\relax}%
\providecommand \@@startlink[1]{}%
\providecommand \@@endlink[0]{}%
\providecommand \url  [0]{\begingroup\@sanitize@url \@url }%
\providecommand \@url [1]{\endgroup\@href {#1}{\urlprefix }}%
\providecommand \urlprefix  [0]{URL }%
\providecommand \Eprint [0]{\href }%
\providecommand \doibase [0]{http://dx.doi.org/}%
\providecommand \selectlanguage [0]{\@gobble}%
\providecommand \bibinfo  [0]{\@secondoftwo}%
\providecommand \bibfield  [0]{\@secondoftwo}%
\providecommand \translation [1]{[#1]}%
\providecommand \BibitemOpen [0]{}%
\providecommand \bibitemStop [0]{}%
\providecommand \bibitemNoStop [0]{.\EOS\space}%
\providecommand \EOS [0]{\spacefactor3000\relax}%
\providecommand \BibitemShut  [1]{\csname bibitem#1\endcsname}%
\let\auto@bib@innerbib\@empty
\bibitem [{\citenamefont {Szulczewski}\ \emph {et~al.}(2012)\citenamefont
  {Szulczewski}, \citenamefont {MacMinn}, \citenamefont {Herzog},\ and\
  \citenamefont {Juanes}}]{juanes12}%
  \BibitemOpen
  \bibfield  {author} {\bibinfo {author} {\bibfnamefont {M.~L.}\ \bibnamefont
  {Szulczewski}}, \bibinfo {author} {\bibfnamefont {C.~W.}\ \bibnamefont
  {MacMinn}}, \bibinfo {author} {\bibfnamefont {H.~J.}\ \bibnamefont {Herzog}},
  \ and\ \bibinfo {author} {\bibfnamefont {R.}~\bibnamefont {Juanes}},\ }\href
  {\doibase 10.1073/pnas.1115347109} {\bibfield  {journal} {\bibinfo  {journal}
  {Proceedings of the National Academy of Sciences}\ }\textbf {\bibinfo
  {volume} {109}},\ \bibinfo {pages} {5185} (\bibinfo {year}
  {2012})}\BibitemShut {NoStop}%
\bibitem [{\citenamefont {Orr}\ and\ \citenamefont {Taber}(1984)}]{orr84}%
  \BibitemOpen
  \bibfield  {author} {\bibinfo {author} {\bibfnamefont {F.}~\bibnamefont
  {Orr}}\ and\ \bibinfo {author} {\bibfnamefont {J.}~\bibnamefont {Taber}},\
  }\href {\doibase 10.1126/science.224.4649.563} {\bibfield  {journal}
  {\bibinfo  {journal} {Science}\ }\textbf {\bibinfo {volume} {224}},\ \bibinfo
  {pages} {563} (\bibinfo {year} {1984})}\BibitemShut {NoStop}%
\bibitem [{\citenamefont {Cueto-Felgueroso}\ and\ \citenamefont
  {Juanes}(2008)}]{juanes08}%
  \BibitemOpen
  \bibfield  {author} {\bibinfo {author} {\bibfnamefont {L.}~\bibnamefont
  {Cueto-Felgueroso}}\ and\ \bibinfo {author} {\bibfnamefont {R.}~\bibnamefont
  {Juanes}},\ }\href {\doibase 10.1103/PhysRevLett.101.244504} {\bibfield
  {journal} {\bibinfo  {journal} {Physical Review Letters}\ }\textbf {\bibinfo
  {volume} {101}},\ \bibinfo {pages} {244504} (\bibinfo {year}
  {2008})}\BibitemShut {NoStop}%
\bibitem [{\citenamefont {Zamani}\ and\ \citenamefont
  {Maini}(2009)}]{zamani09}%
  \BibitemOpen
  \bibfield  {author} {\bibinfo {author} {\bibfnamefont {A.}~\bibnamefont
  {Zamani}}\ and\ \bibinfo {author} {\bibfnamefont {B.}~\bibnamefont {Maini}},\
  }\href {\doibase 10.1016/j.petrol.2009.06.016} {\bibfield  {journal}
  {\bibinfo  {journal} {Journal of Petroleum Science and Engineering}\ }\textbf
  {\bibinfo {volume} {69}},\ \bibinfo {pages} {71} (\bibinfo {year}
  {2009})}\BibitemShut {NoStop}%
\bibitem [{\citenamefont {Stefanescu}(2002)}]{stefanescu02}%
  \BibitemOpen
  \bibfield  {author} {\bibinfo {author} {\bibfnamefont {D.~M.}\ \bibnamefont
  {Stefanescu}},\ }\href {\doibase 10.1007/978-3-319-15693-4} {\emph {\bibinfo
  {title} {Science and Engineering of Casting Solidification}}}\ (\bibinfo
  {publisher} {Springer},\ \bibinfo {year} {2002})\ pp.\ \bibinfo {pages}
  {49--46}\BibitemShut {NoStop}%
\bibitem [{\citenamefont {Khaled}\ and\ \citenamefont
  {Vafai}(2003)}]{khaled03}%
  \BibitemOpen
  \bibfield  {author} {\bibinfo {author} {\bibfnamefont {A.-R.}\ \bibnamefont
  {Khaled}}\ and\ \bibinfo {author} {\bibfnamefont {K.}~\bibnamefont {Vafai}},\
  }\href {\doibase https://doi.org/10.1016/S0017-9310(03)00301-6} {\bibfield
  {journal} {\bibinfo  {journal} {International Journal of Heat and Mass
  Transfer}\ }\textbf {\bibinfo {volume} {46}},\ \bibinfo {pages} {4989 }
  (\bibinfo {year} {2003})}\BibitemShut {NoStop}%
\bibitem [{\citenamefont {Delgado}(2011)}]{miguel12}%
  \BibitemOpen
  \bibfield  {author} {\bibinfo {author} {\bibfnamefont {J.~M.}\ \bibnamefont
  {Delgado}},\ }\href {\doibase 10.1007/978-3-642-21966-5} {\emph {\bibinfo
  {title} {Heat and mass transfer in porous media}}},\ Vol.~\bibinfo {volume}
  {13}\ (\bibinfo  {publisher} {Springer},\ \bibinfo {year} {2011})\ pp.\
  \bibinfo {pages} {115--137}\BibitemShut {NoStop}%
\bibitem [{\citenamefont {Linninger}\ \emph {et~al.}(2007)\citenamefont
  {Linninger}, \citenamefont {Xenos}, \citenamefont {Zhu}, \citenamefont
  {Somayaji}, \citenamefont {Kondapalli},\ and\ \citenamefont {Penn}}]{penn07}%
  \BibitemOpen
  \bibfield  {author} {\bibinfo {author} {\bibfnamefont {A.~A.}\ \bibnamefont
  {Linninger}}, \bibinfo {author} {\bibfnamefont {M.}~\bibnamefont {Xenos}},
  \bibinfo {author} {\bibfnamefont {D.~C.}\ \bibnamefont {Zhu}}, \bibinfo
  {author} {\bibfnamefont {M.~R.}\ \bibnamefont {Somayaji}}, \bibinfo {author}
  {\bibfnamefont {S.}~\bibnamefont {Kondapalli}}, \ and\ \bibinfo {author}
  {\bibfnamefont {R.~D.}\ \bibnamefont {Penn}},\ }\href {\doibase
  10.1109/TBME.2006.886853} {\bibfield  {journal} {\bibinfo  {journal} {IEEE
  Transactions on Biomedical Engineering}\ }\textbf {\bibinfo {volume} {54}},\
  \bibinfo {pages} {291} (\bibinfo {year} {2007})}\BibitemShut {NoStop}%
\bibitem [{\citenamefont {D'Angelo}\ and\ \citenamefont
  {Zunino}(2011)}]{zunino10}%
  \BibitemOpen
  \bibfield  {author} {\bibinfo {author} {\bibfnamefont {C.}~\bibnamefont
  {D'Angelo}}\ and\ \bibinfo {author} {\bibfnamefont {P.}~\bibnamefont
  {Zunino}},\ }\href {\doibase 10.1051/m2an/2010062} {\bibfield  {journal}
  {\bibinfo  {journal} {ESAIM: Mathematical Modelling and Numerical Analysis}\
  }\textbf {\bibinfo {volume} {45}},\ \bibinfo {pages} {447} (\bibinfo {year}
  {2011})}\BibitemShut {NoStop}%
\bibitem [{\citenamefont {Andrade~Jr}\ \emph {et~al.}(1999)\citenamefont
  {Andrade~Jr}, \citenamefont {Costa}, \citenamefont {Almeida}, \citenamefont
  {Makse},\ and\ \citenamefont {Stanley}}]{stanley99}%
  \BibitemOpen
  \bibfield  {author} {\bibinfo {author} {\bibfnamefont {J.}~\bibnamefont
  {Andrade~Jr}}, \bibinfo {author} {\bibfnamefont {U.}~\bibnamefont {Costa}},
  \bibinfo {author} {\bibfnamefont {M.}~\bibnamefont {Almeida}}, \bibinfo
  {author} {\bibfnamefont {H.}~\bibnamefont {Makse}}, \ and\ \bibinfo {author}
  {\bibfnamefont {H.}~\bibnamefont {Stanley}},\ }\href {\doibase
  10.1103/PhysRevLett.82.5249} {\bibfield  {journal} {\bibinfo  {journal}
  {Physical Review Letters}\ }\textbf {\bibinfo {volume} {82}},\ \bibinfo
  {pages} {5249} (\bibinfo {year} {1999})}\BibitemShut {NoStop}%
\bibitem [{\citenamefont {Chen}\ and\ \citenamefont {Doolen}(1998)}]{chen98}%
  \BibitemOpen
  \bibfield  {author} {\bibinfo {author} {\bibfnamefont {S.}~\bibnamefont
  {Chen}}\ and\ \bibinfo {author} {\bibfnamefont {G.~D.}\ \bibnamefont
  {Doolen}},\ }\href {\doibase 10.1146/annurev.fluid.30.1.329} {\bibfield
  {journal} {\bibinfo  {journal} {Annual review of fluid mechanics}\ }\textbf
  {\bibinfo {volume} {30}},\ \bibinfo {pages} {329} (\bibinfo {year}
  {1998})}\BibitemShut {NoStop}%
\bibitem [{\citenamefont {Rothman}(1988)}]{rothman88}%
  \BibitemOpen
  \bibfield  {author} {\bibinfo {author} {\bibfnamefont {D.~H.}\ \bibnamefont
  {Rothman}},\ }\href {\doibase 10.1190/1.1442482} {\bibfield  {journal}
  {\bibinfo  {journal} {Geophysics}\ }\textbf {\bibinfo {volume} {53}},\
  \bibinfo {pages} {509} (\bibinfo {year} {1988})}\BibitemShut {NoStop}%
\bibitem [{\citenamefont {Hunt}\ \emph {et~al.}(2014)\citenamefont {Hunt},
  \citenamefont {Ewing},\ and\ \citenamefont {Ghanbarian}}]{hunt14}%
  \BibitemOpen
  \bibfield  {author} {\bibinfo {author} {\bibfnamefont {A.}~\bibnamefont
  {Hunt}}, \bibinfo {author} {\bibfnamefont {R.}~\bibnamefont {Ewing}}, \ and\
  \bibinfo {author} {\bibfnamefont {B.}~\bibnamefont {Ghanbarian}},\ }\href
  {\doibase 10.1007/978-3-319-03771-4} {\emph {\bibinfo {title} {Percolation
  theory for flow in porous media}}},\ Vol.\ \bibinfo {volume} {880}\ (\bibinfo
   {publisher} {Springer},\ \bibinfo {year} {2014})\BibitemShut {NoStop}%
\bibitem [{\citenamefont {King}\ and\ \citenamefont {Masihi}(2018)}]{king18}%
  \BibitemOpen
  \bibfield  {author} {\bibinfo {author} {\bibfnamefont {P.~R.}\ \bibnamefont
  {King}}\ and\ \bibinfo {author} {\bibfnamefont {M.}~\bibnamefont {Masihi}},\
  }\href {\doibase 10.1142/q0154} {\emph {\bibinfo {title} {Percolation Theory
  in Reservoir Engineering}}}\ (\bibinfo  {publisher} {World Scientific},\
  \bibinfo {year} {2018})\BibitemShut {NoStop}%
\bibitem [{\citenamefont {King}(1987)}]{king87}%
  \BibitemOpen
  \bibfield  {author} {\bibinfo {author} {\bibfnamefont {P.}~\bibnamefont
  {King}},\ }\href {\doibase 10.1088/0305-4470/20/12/038} {\bibfield  {journal}
  {\bibinfo  {journal} {Journal of Physics A: Mathematical and General}\
  }\textbf {\bibinfo {volume} {20}},\ \bibinfo {pages} {3935} (\bibinfo {year}
  {1987})}\BibitemShut {NoStop}%
\bibitem [{\citenamefont {Sahimi}(1993)}]{sahimi93}%
  \BibitemOpen
  \bibfield  {author} {\bibinfo {author} {\bibfnamefont {M.}~\bibnamefont
  {Sahimi}},\ }\href {\doibase 10.1103/RevModPhys.65.1393} {\bibfield
  {journal} {\bibinfo  {journal} {Reviews of modern physics}\ }\textbf
  {\bibinfo {volume} {65}},\ \bibinfo {pages} {1393} (\bibinfo {year}
  {1993})}\BibitemShut {NoStop}%
\bibitem [{\citenamefont {Blunt}(2017)}]{blunt17}%
  \BibitemOpen
  \bibfield  {author} {\bibinfo {author} {\bibfnamefont {M.~J.}\ \bibnamefont
  {Blunt}},\ }\href {\doibase 10.1017/9781316145098} {\emph {\bibinfo {title}
  {Multiphase flow in permeable media: A pore-scale perspective}}}\ (\bibinfo
  {publisher} {Cambridge University Press},\ \bibinfo {year} {2017})\ pp.\
  \bibinfo {pages} {77--103}\BibitemShut {NoStop}%
\bibitem [{\citenamefont {King}(1989)}]{king89}%
  \BibitemOpen
  \bibfield  {author} {\bibinfo {author} {\bibfnamefont {P.~R.}\ \bibnamefont
  {King}},\ }\href {\doibase 10.1007/BF00134741} {\bibfield  {journal}
  {\bibinfo  {journal} {Transport in Porous Media}\ }\textbf {\bibinfo {volume}
  {4}},\ \bibinfo {pages} {37} (\bibinfo {year} {1989})}\BibitemShut {NoStop}%
\bibitem [{\citenamefont {Morris}\ and\ \citenamefont {Ball}(1990)}]{morris90}%
  \BibitemOpen
  \bibfield  {author} {\bibinfo {author} {\bibfnamefont {M.}~\bibnamefont
  {Morris}}\ and\ \bibinfo {author} {\bibfnamefont {R.}~\bibnamefont {Ball}},\
  }\href {\doibase 10.1088/0305-4470/23/19/009} {\bibfield  {journal} {\bibinfo
   {journal} {Journal of Physics A: Mathematical and General}\ }\textbf
  {\bibinfo {volume} {23}},\ \bibinfo {pages} {4199} (\bibinfo {year}
  {1990})}\BibitemShut {NoStop}%
\bibitem [{\citenamefont {King}\ \emph {et~al.}(1993)\citenamefont {King},
  \citenamefont {Muggeridge},\ and\ \citenamefont {Price}}]{king93}%
  \BibitemOpen
  \bibfield  {author} {\bibinfo {author} {\bibfnamefont {P.}~\bibnamefont
  {King}}, \bibinfo {author} {\bibfnamefont {A.}~\bibnamefont {Muggeridge}}, \
  and\ \bibinfo {author} {\bibfnamefont {W.}~\bibnamefont {Price}},\ }\href
  {\doibase 10.1007/BF00624460} {\bibfield  {journal} {\bibinfo  {journal}
  {Transport in Porous Media}\ }\textbf {\bibinfo {volume} {12}},\ \bibinfo
  {pages} {237} (\bibinfo {year} {1993})}\BibitemShut {NoStop}%
\bibitem [{\citenamefont {King}\ and\ \citenamefont
  {Neuweiler}(2002)}]{king02}%
  \BibitemOpen
  \bibfield  {author} {\bibinfo {author} {\bibfnamefont {P.}~\bibnamefont
  {King}}\ and\ \bibinfo {author} {\bibfnamefont {I.}~\bibnamefont
  {Neuweiler}},\ }\href {\doibase 10.1023/A:1016533230647} {\bibfield
  {journal} {\bibinfo  {journal} {Computational Geosciences}\ }\textbf
  {\bibinfo {volume} {6}},\ \bibinfo {pages} {101} (\bibinfo {year}
  {2002})}\BibitemShut {NoStop}%
\bibitem [{\citenamefont {Neuweiler}\ \emph {et~al.}(2003)\citenamefont
  {Neuweiler}, \citenamefont {Attinger}, \citenamefont {Kinzelbach},\ and\
  \citenamefont {King}}]{neuweiler03}%
  \BibitemOpen
  \bibfield  {author} {\bibinfo {author} {\bibfnamefont {I.}~\bibnamefont
  {Neuweiler}}, \bibinfo {author} {\bibfnamefont {S.}~\bibnamefont {Attinger}},
  \bibinfo {author} {\bibfnamefont {W.}~\bibnamefont {Kinzelbach}}, \ and\
  \bibinfo {author} {\bibfnamefont {P.}~\bibnamefont {King}},\ }\href {\doibase
  10.1023/A:1022370927468} {\bibfield  {journal} {\bibinfo  {journal}
  {Transport in Porous Media}\ }\textbf {\bibinfo {volume} {51}},\ \bibinfo
  {pages} {287} (\bibinfo {year} {2003})}\BibitemShut {NoStop}%
\bibitem [{\citenamefont {Pancaldi}\ \emph {et~al.}(2007)\citenamefont
  {Pancaldi}, \citenamefont {Christensen},\ and\ \citenamefont
  {King}}]{pancaldi07}%
  \BibitemOpen
  \bibfield  {author} {\bibinfo {author} {\bibfnamefont {V.}~\bibnamefont
  {Pancaldi}}, \bibinfo {author} {\bibfnamefont {K.}~\bibnamefont
  {Christensen}}, \ and\ \bibinfo {author} {\bibfnamefont {P.~R.}\ \bibnamefont
  {King}},\ }\href {\doibase 10.1007/s11242-006-9032-0} {\bibfield  {journal}
  {\bibinfo  {journal} {Transport in porous media}\ }\textbf {\bibinfo {volume}
  {67}},\ \bibinfo {pages} {395} (\bibinfo {year} {2007})}\BibitemShut
  {NoStop}%
\bibitem [{\citenamefont {Pancaldi}\ \emph {et~al.}(2008)\citenamefont
  {Pancaldi}, \citenamefont {King},\ and\ \citenamefont
  {Christensen}}]{pancaldi08}%
  \BibitemOpen
  \bibfield  {author} {\bibinfo {author} {\bibfnamefont {V.}~\bibnamefont
  {Pancaldi}}, \bibinfo {author} {\bibfnamefont {P.~R.}\ \bibnamefont {King}},
  \ and\ \bibinfo {author} {\bibfnamefont {K.}~\bibnamefont {Christensen}},\
  }\href {\doibase 10.1016/j.physa.2008.03.031} {\bibfield  {journal} {\bibinfo
   {journal} {Physica A: Statistical Mechanics and its Applications}\ }\textbf
  {\bibinfo {volume} {387}},\ \bibinfo {pages} {4760} (\bibinfo {year}
  {2008})}\BibitemShut {NoStop}%
\bibitem [{\citenamefont {Pancaldi}\ \emph {et~al.}(2009)\citenamefont
  {Pancaldi}, \citenamefont {King},\ and\ \citenamefont
  {Christensen}}]{pancaldi09}%
  \BibitemOpen
  \bibfield  {author} {\bibinfo {author} {\bibfnamefont {V.}~\bibnamefont
  {Pancaldi}}, \bibinfo {author} {\bibfnamefont {P.~R.}\ \bibnamefont {King}},
  \ and\ \bibinfo {author} {\bibfnamefont {K.}~\bibnamefont {Christensen}},\
  }\href {\doibase 10.1103/PhysRevE.79.036704} {\bibfield  {journal} {\bibinfo
  {journal} {Physical Review E}\ }\textbf {\bibinfo {volume} {79}},\ \bibinfo
  {pages} {036704} (\bibinfo {year} {2009})}\BibitemShut {NoStop}%
\bibitem [{\citenamefont {Teodorovich}(1997)}]{teodorovich97}%
  \BibitemOpen
  \bibfield  {author} {\bibinfo {author} {\bibfnamefont {E.}~\bibnamefont
  {Teodorovich}},\ }\href {\doibase 10.1134/1.558302} {\bibfield  {journal}
  {\bibinfo  {journal} {Journal of Experimental and Theoretical Physics}\
  }\textbf {\bibinfo {volume} {85}},\ \bibinfo {pages} {173} (\bibinfo {year}
  {1997})}\BibitemShut {NoStop}%
\bibitem [{\citenamefont {Hristopulos}\ and\ \citenamefont
  {Christakos}(1999)}]{hristopulos99}%
  \BibitemOpen
  \bibfield  {author} {\bibinfo {author} {\bibfnamefont {D.}~\bibnamefont
  {Hristopulos}}\ and\ \bibinfo {author} {\bibfnamefont {G.}~\bibnamefont
  {Christakos}},\ }\href {\doibase 10.1007/s004770050036} {\bibfield  {journal}
  {\bibinfo  {journal} {Stochastic Environmental Research and Risk Assessment}\
  }\textbf {\bibinfo {volume} {13}},\ \bibinfo {pages} {131} (\bibinfo {year}
  {1999})}\BibitemShut {NoStop}%
\bibitem [{\citenamefont {Sposito}(2001)}]{sposito01}%
  \BibitemOpen
  \bibfield  {author} {\bibinfo {author} {\bibfnamefont {G.}~\bibnamefont
  {Sposito}},\ }\href {\doibase 10.1023/A:1006724801223} {\bibfield  {journal}
  {\bibinfo  {journal} {Transport in Porous Media}\ }\textbf {\bibinfo {volume}
  {42}},\ \bibinfo {pages} {181} (\bibinfo {year} {2001})}\BibitemShut
  {NoStop}%
\bibitem [{\citenamefont {Attinger}(2003)}]{attinger03}%
  \BibitemOpen
  \bibfield  {author} {\bibinfo {author} {\bibfnamefont {S.}~\bibnamefont
  {Attinger}},\ }\href {\doibase 10.1023/B:COMG.0000005243.73381.e3} {\bibfield
   {journal} {\bibinfo  {journal} {Computational Geosciences}\ }\textbf
  {\bibinfo {volume} {7}},\ \bibinfo {pages} {253} (\bibinfo {year}
  {2003})}\BibitemShut {NoStop}%
\bibitem [{\citenamefont {Eberhard}\ \emph {et~al.}(2004)\citenamefont
  {Eberhard}, \citenamefont {Attinger},\ and\ \citenamefont
  {Wittum}}]{eberhard04}%
  \BibitemOpen
  \bibfield  {author} {\bibinfo {author} {\bibfnamefont {J.}~\bibnamefont
  {Eberhard}}, \bibinfo {author} {\bibfnamefont {S.}~\bibnamefont {Attinger}},
  \ and\ \bibinfo {author} {\bibfnamefont {G.}~\bibnamefont {Wittum}},\ }\href
  {\doibase doi.org/10.1137/030600497} {\bibfield  {journal} {\bibinfo
  {journal} {Multiscale Modeling \& Simulation}\ }\textbf {\bibinfo {volume}
  {2}},\ \bibinfo {pages} {269} (\bibinfo {year} {2004})}\BibitemShut {NoStop}%
\bibitem [{\citenamefont {Hanasoge}\ \emph {et~al.}(2017)\citenamefont
  {Hanasoge}, \citenamefont {Agarwal}, \citenamefont {Tandon},\ and\
  \citenamefont {Koelman}}]{hanasoge17}%
  \BibitemOpen
  \bibfield  {author} {\bibinfo {author} {\bibfnamefont {S.}~\bibnamefont
  {Hanasoge}}, \bibinfo {author} {\bibfnamefont {U.}~\bibnamefont {Agarwal}},
  \bibinfo {author} {\bibfnamefont {K.}~\bibnamefont {Tandon}}, \ and\ \bibinfo
  {author} {\bibfnamefont {J.~V.~A.}\ \bibnamefont {Koelman}},\ }\href
  {\doibase 10.1103/PhysRevE.96.033313} {\bibfield  {journal} {\bibinfo
  {journal} {Physical Review E}\ }\textbf {\bibinfo {volume} {96}},\ \bibinfo
  {pages} {033313} (\bibinfo {year} {2017})}\BibitemShut {NoStop}%
\bibitem [{\citenamefont {Westbroek}\ \emph
  {et~al.}(2019{\natexlab{a}})\citenamefont {Westbroek}, \citenamefont {Coche},
  \citenamefont {King},\ and\ \citenamefont {Vvedensky}}]{westbroek19a}%
  \BibitemOpen
  \bibfield  {author} {\bibinfo {author} {\bibfnamefont {M.~J.}\ \bibnamefont
  {Westbroek}}, \bibinfo {author} {\bibfnamefont {G.-A.}\ \bibnamefont
  {Coche}}, \bibinfo {author} {\bibfnamefont {P.~R.}\ \bibnamefont {King}}, \
  and\ \bibinfo {author} {\bibfnamefont {D.~D.}\ \bibnamefont {Vvedensky}},\
  }\href {\doibase 10.1088/1751-8121/ab1100} {\bibfield  {journal} {\bibinfo
  {journal} {Journal of Physics A: Mathematical and Theoretical}\ }\textbf
  {\bibinfo {volume} {52}},\ \bibinfo {pages} {185001} (\bibinfo {year}
  {2019}{\natexlab{a}})}\BibitemShut {NoStop}%
\bibitem [{\citenamefont {Westbroek}\ \emph
  {et~al.}(2019{\natexlab{b}})\citenamefont {Westbroek}, \citenamefont {King},
  \citenamefont {Vvedensky},\ and\ \citenamefont {Schwede}}]{westbroek19b}%
  \BibitemOpen
  \bibfield  {author} {\bibinfo {author} {\bibfnamefont {M.~J.~E.}\
  \bibnamefont {Westbroek}}, \bibinfo {author} {\bibfnamefont {P.~R.}\
  \bibnamefont {King}}, \bibinfo {author} {\bibfnamefont {D.~D.}\ \bibnamefont
  {Vvedensky}}, \ and\ \bibinfo {author} {\bibfnamefont {R.~L.}\ \bibnamefont
  {Schwede}},\ }\href@noop {} {\enquote {\bibinfo {title} {Pressure and flow
  statistics of darcy flow from simulated annealing},}\ } (\bibinfo {year}
  {2019}{\natexlab{b}}),\ \Eprint {http://arxiv.org/abs/1903.10439}
  {arXiv:1903.10439 [cs.CE]} \BibitemShut {NoStop}%
\bibitem [{\citenamefont {Darcy}(1856)}]{darcy56}%
  \BibitemOpen
  \bibfield  {author} {\bibinfo {author} {\bibfnamefont {H.~P.~G.}\
  \bibnamefont {Darcy}},\ }\href@noop {} {\emph {\bibinfo {title} {Les
  Fontaines publiques de la ville de Dijon. Exposition et application des
  principes {\`a} suivre et des formules {\`a} employer dans les questions de
  distribution d'eau, etc}}}\ (\bibinfo  {publisher} {V. Dalamont},\ \bibinfo
  {year} {1856})\BibitemShut {NoStop}%
\bibitem [{\citenamefont {Whitaker}(1986)}]{whitaker86}%
  \BibitemOpen
  \bibfield  {author} {\bibinfo {author} {\bibfnamefont {S.}~\bibnamefont
  {Whitaker}},\ }\href {\doibase 10.1007/BF01376989} {\bibfield  {journal}
  {\bibinfo  {journal} {Transport in porous media}\ }\textbf {\bibinfo {volume}
  {1}},\ \bibinfo {pages} {3} (\bibinfo {year} {1986})}\BibitemShut {NoStop}%
\bibitem [{\citenamefont {Rubinstein}(1987)}]{rubinstein87}%
  \BibitemOpen
  \bibfield  {author} {\bibinfo {author} {\bibfnamefont {J.}~\bibnamefont
  {Rubinstein}},\ }in\ \href {\doibase 10.1007/978-1-4684-6347-7_12} {\emph
  {\bibinfo {booktitle} {Hydrodynamic behavior and interacting particle
  systems}}}\ (\bibinfo  {publisher} {Springer},\ \bibinfo {year} {1987})\ pp.\
  \bibinfo {pages} {137--149}\BibitemShut {NoStop}%
\bibitem [{\citenamefont {De~Dominicis}\ and\ \citenamefont
  {Peliti}(1978)}]{dedominicis78}%
  \BibitemOpen
  \bibfield  {author} {\bibinfo {author} {\bibfnamefont {C.}~\bibnamefont
  {De~Dominicis}}\ and\ \bibinfo {author} {\bibfnamefont {L.}~\bibnamefont
  {Peliti}},\ }\href {\doibase 10.1103/PhysRevB.18.353} {\bibfield  {journal}
  {\bibinfo  {journal} {Physical Review B}\ }\textbf {\bibinfo {volume} {18}},\
  \bibinfo {pages} {353} (\bibinfo {year} {1978})}\BibitemShut {NoStop}%
\bibitem [{\citenamefont {Wilson}\ and\ \citenamefont
  {Kogut}(1974)}]{wilson74}%
  \BibitemOpen
  \bibfield  {author} {\bibinfo {author} {\bibfnamefont {K.~G.}\ \bibnamefont
  {Wilson}}\ and\ \bibinfo {author} {\bibfnamefont {J.}~\bibnamefont {Kogut}},\
  }\href {\doibase 10.1016/0370-1573(74)90023-4} {\bibfield  {journal}
  {\bibinfo  {journal} {Physics reports}\ }\textbf {\bibinfo {volume} {12}},\
  \bibinfo {pages} {75} (\bibinfo {year} {1974})}\BibitemShut {NoStop}%
\bibitem [{\citenamefont {Goldenfeld}(1992)}]{goldenfeld92}%
  \BibitemOpen
  \bibfield  {author} {\bibinfo {author} {\bibfnamefont {N.}~\bibnamefont
  {Goldenfeld}},\ }\href {\doibase 10.1201/9780429493492} {\emph {\bibinfo
  {title} {Lectures on phase transitions and the renormalization group}}}\
  (\bibinfo  {publisher} {CRC Press},\ \bibinfo {year} {1992})\BibitemShut
  {NoStop}%
\bibitem [{\citenamefont {Chang}\ \emph {et~al.}(1992)\citenamefont {Chang},
  \citenamefont {Vvedensky},\ and\ \citenamefont {Nicoll}}]{chang92}%
  \BibitemOpen
  \bibfield  {author} {\bibinfo {author} {\bibfnamefont {T.~S.}\ \bibnamefont
  {Chang}}, \bibinfo {author} {\bibfnamefont {D.~D.}\ \bibnamefont
  {Vvedensky}}, \ and\ \bibinfo {author} {\bibfnamefont {J.~F.}\ \bibnamefont
  {Nicoll}},\ }\href {\doibase 10.1016/0370-1573(92)90041-W} {\bibfield
  {journal} {\bibinfo  {journal} {Physics Reports}\ }\textbf {\bibinfo {volume}
  {217}},\ \bibinfo {pages} {279} (\bibinfo {year} {1992})}\BibitemShut
  {NoStop}%
\bibitem [{Note1()}]{Note1}%
  \BibitemOpen
  \bibinfo {note} {For Gaussian random variables, this result was established
  by Isserlis \cite {isserlis18}. In the context of quantum field theory, the
  analogous result is Wick's theorem \cite {wick50}.}\BibitemShut {Stop}%
\bibitem [{\citenamefont {Täuber}(2014)}]{tauber14}%
  \BibitemOpen
  \bibfield  {author} {\bibinfo {author} {\bibfnamefont {U.~C.}\ \bibnamefont
  {Täuber}},\ }\href {\doibase 10.1017/CBO9781139046213} {\emph {\bibinfo
  {title} {Critical Dynamics: A Field Theory Approach to Equilibrium and
  Non-Equilibrium Scaling Behavior}}}\ (\bibinfo  {publisher} {Cambridge
  University Press},\ \bibinfo {year} {2014})\ p.~\bibinfo {pages}
  {30}\BibitemShut {NoStop}%
\bibitem [{\citenamefont {Deem}\ and\ \citenamefont {Chandler}(1994)}]{deem94}%
  \BibitemOpen
  \bibfield  {author} {\bibinfo {author} {\bibfnamefont {M.~W.}\ \bibnamefont
  {Deem}}\ and\ \bibinfo {author} {\bibfnamefont {D.}~\bibnamefont
  {Chandler}},\ }\href {\doibase 10.1007/BF02188692} {\bibfield  {journal}
  {\bibinfo  {journal} {Journal of statistical physics}\ }\textbf {\bibinfo
  {volume} {76}},\ \bibinfo {pages} {911} (\bibinfo {year} {1994})}\BibitemShut
  {NoStop}%
\bibitem [{\citenamefont {Law}\ \emph {et~al.}(1944)\citenamefont {Law} \emph
  {et~al.}}]{law44}%
  \BibitemOpen
  \bibfield  {author} {\bibinfo {author} {\bibfnamefont {J.}~\bibnamefont
  {Law}} \emph {et~al.},\ }\href {\doibase 10.2118/944202-G} {\bibfield
  {journal} {\bibinfo  {journal} {Transactions of the AIME}\ }\textbf {\bibinfo
  {volume} {155}},\ \bibinfo {pages} {202} (\bibinfo {year}
  {1944})}\BibitemShut {NoStop}%
\bibitem [{\citenamefont {Li}\ \emph {et~al.}(2005)\citenamefont {Li},
  \citenamefont {LeBoeuf}, \citenamefont {Basu},\ and\ \citenamefont
  {Mahadevan}}]{li05}%
  \BibitemOpen
  \bibfield  {author} {\bibinfo {author} {\bibfnamefont {Y.}~\bibnamefont
  {Li}}, \bibinfo {author} {\bibfnamefont {E.~J.}\ \bibnamefont {LeBoeuf}},
  \bibinfo {author} {\bibfnamefont {P.~K.}\ \bibnamefont {Basu}}, \ and\
  \bibinfo {author} {\bibfnamefont {S.}~\bibnamefont {Mahadevan}},\ }\href
  {\doibase 10.1016/j.advwatres.2005.01.007} {\bibfield  {journal} {\bibinfo
  {journal} {Advances in Water Resources}\ }\textbf {\bibinfo {volume} {28}},\
  \bibinfo {pages} {835} (\bibinfo {year} {2005})}\BibitemShut {NoStop}%
\bibitem [{\citenamefont {Phythian}(1977)}]{phythian77}%
  \BibitemOpen
  \bibfield  {author} {\bibinfo {author} {\bibfnamefont {R.}~\bibnamefont
  {Phythian}},\ }\href {\doibase 10.1088/0305-4470/10/5/011} {\bibfield
  {journal} {\bibinfo  {journal} {Journal of Physics A: Mathematical and
  General}\ }\textbf {\bibinfo {volume} {10}},\ \bibinfo {pages} {777}
  (\bibinfo {year} {1977})}\BibitemShut {NoStop}%
\bibitem [{\citenamefont {Buckley}\ \emph {et~al.}(1942)\citenamefont
  {Buckley}, \citenamefont {Leverett} \emph {et~al.}}]{buckley42}%
  \BibitemOpen
  \bibfield  {author} {\bibinfo {author} {\bibfnamefont {S.~E.}\ \bibnamefont
  {Buckley}}, \bibinfo {author} {\bibfnamefont {M.}~\bibnamefont {Leverett}},
  \emph {et~al.},\ }\href {\doibase 10.2118/942107-G} {\bibfield  {journal}
  {\bibinfo  {journal} {Transactions of the AIME}\ }\textbf {\bibinfo {volume}
  {146}},\ \bibinfo {pages} {107} (\bibinfo {year} {1942})}\BibitemShut
  {NoStop}%
\bibitem [{\citenamefont {Isserlis}(1918)}]{isserlis18}%
  \BibitemOpen
  \bibfield  {author} {\bibinfo {author} {\bibfnamefont {L.}~\bibnamefont
  {Isserlis}},\ }\href {\doibase 10.2307/2331932} {\bibfield  {journal}
  {\bibinfo  {journal} {Biometrika}\ }\textbf {\bibinfo {volume} {12}},\
  \bibinfo {pages} {134} (\bibinfo {year} {1918})}\BibitemShut {NoStop}%
\bibitem [{\citenamefont {Wick}(1950)}]{wick50}%
  \BibitemOpen
  \bibfield  {author} {\bibinfo {author} {\bibfnamefont {G.-C.}\ \bibnamefont
  {Wick}},\ }\href {\doibase 10.1103/PhysRev.80.268} {\bibfield  {journal}
  {\bibinfo  {journal} {Physical review}\ }\textbf {\bibinfo {volume} {80}},\
  \bibinfo {pages} {268} (\bibinfo {year} {1950})}\BibitemShut {NoStop}%
\end{thebibliography}%

\end{document}